\documentclass[aps,prd,twocolumn,showpacs,amsmath,amssymb,nofootinbib,eqsecnum, preprintnumbers]{revtex4-1}

\usepackage{graphicx}

\usepackage{epstopdf}
\usepackage{color}

\newcommand{\be}{\begin{equation}}
\newcommand{\ee}{\end{equation}}
\newcommand{\ba}{\begin{eqnarray}}
\newcommand{\ea}{\end{eqnarray}}

\newcommand{\beq}{\begin{equation}}
\newcommand{\eeq}{\end{equation}}
\newcommand{\beqa}{\begin{eqnarray}}
\newcommand{\eeqa}{\end{eqnarray}}

\begin{document}
\title{Extended phase space  thermodynamics for charged and rotating black holes and 
Born--Infeld vacuum polarization}

\author{Sharmila Gunasekaran}
\email{sharmila.dhevi@gmail.com}
\affiliation{Department of Physics and Astronomy, University of Waterloo,
Waterloo, Ontario, Canada, N2L 3G1}

\author{David Kubiz\v n\'ak}
\email{dkubiznak@perimeterinstitute.ca}
\affiliation{Perimeter Institute, 31 Caroline St. N. Waterloo
Ontario, N2L 2Y5, Canada}

\author{Robert B. Mann}
\email{rbmann@sciborg.uwaterloo.ca}
\affiliation{Department of Physics and Astronomy, University of Waterloo,
Waterloo, Ontario, Canada, N2L 3G1}

\date{November 15, 2012}  

\begin{abstract}{
We investigate the critical behaviour of charged and rotating AdS black holes  in $d$ spacetime dimensions, including effects
from non-linear electrodynamics via the Born-Infeld action, in an extended phase space in which the cosmological constant is interpreted as thermodynamic pressure. 
 For Reissner--N\"ordstrom black holes
we find that the analogy with the Van der Walls liquid--gas system holds in any dimension greater than three, and that
the critical exponents coincide with those of the Van der Waals system. We find that neutral slowly rotating black holes
in four space-time dimensions also have the same qualitative behaviour. However  charged and rotating black holes
in three spacetime dimensions do not exhibit critical phenomena. 
For Born-Infeld black holes we define a new thermodynamic quantity ${\cal B}$ conjugate to the Born-Infeld parameter $b$
 that we call Born--Infeld vacuum polarization.  We demonstrate that this quantity is required for consistency of both
 the  first law of thermodynamics and  the corresponding Smarr relation. }
\end{abstract}

\preprint{pi-stronggrv-291}

\maketitle

\section{Introduction}

The study of black hole thermodynamics has come a long way since the seminal start by Hawking and Page \cite{HawkingPage:1983}, who demonstrated the existence of a certain phase transition in the phase space of a Schwarzchild-AdS black hole.  Specific
interest developed in the thermodynamics of charged black holes in asymptotically AdS spacetimes,
 in large part  because they admit a gauge duality description via a dual thermal field theory.   This duality description suggested that  
Reissner-N\"ordstrom AdS black holes exhibit critical behaviour suggestive of a Van der Waals liquid gas phase transition \cite{ChamblinEtal:1999a,ChamblinEtal:1999b}.

Recently this picture has been substantively revised  \cite{KubiznakMann:2012}. By treating the cosmological constant as a thermodynamic pressure and its conjugate quantity as a thermodynamic volume, the analogy between 4-dimensional 
Reissner-N\"ordstrom AdS black holes and the Van der Waals liquid--gas system can be completed,  with the  critical exponents coinciding with those of the Van der Waals system and predicted by the mean field theory.  While previous studies considered the conventional phase space of a black hole to consist only of entropy, temperature, charge and potential, investigations of the critical behaviour in the backdrop of an extended phase-space 
(including pressure and volume) seems more meaningful for the following reasons.
\begin{itemize}
\item With the extended phase space, the Smarr relation is satisfied in addition to the first law of thermodynamics, from which it is derived from scaling arguments. This makes it evident that the phase space that should be considered is not the  conventional one  \cite{KastorEtal:2009}.
\item The resulting equation of state can be used for comparison with real world thermodynamic systems \cite{KastorEtal:2009, Dolan:2010, Dolan:2011a, Dolan:2011b, KubiznakMann:2012}.  For example, the variation of these parameters makes it possible to identify the mass of the black hole as enthalpy rather than internal energy.  As a consequence, this may be useful in interpreting the kind of critical behaviour (if any) to the known thermodynamic systems (if any).
\item  Thermodynamic volume has been studied for a wide variety of black holes and is conjectured to satisfy the {\em reverse isoperimetric inequality} \cite{CveticEtal:2010}.
\item  Variation of parameters in the extended phase space turns out to be useful for yet another reason: one can consider more fundamental theories that admit the variation of physical constants  \cite{KastorEtal:2009, CveticEtal:2010}. 
\end{itemize}

As noted above, a recent investigation of 4-dimensional charged (Reissner--N\"ordstrom) AdS black hole 
thermodynamics in an extended phase space---where the cosmological constant is treated as a dynamical pressure and the corresponding conjugate quantity as volume---indicated that the
analogy with a Van der Waals fluid becomes very precise and complete \cite{KubiznakMann:2012}. One can compare appropriately analogous physical quantities, the phase diagrams become extremely similar, and the critical behaviour at the point of second order transition identical. 

In this paper we elaborate on this study by considering more general classes of charged and rotating black holes:   specifically, the
d-dimensional analogues of the Reissner--N\"ordstrom-AdS black hole, its nonlinear generalization described by  Born--Infeld theory,
and the Kerr-Newman solution.   We find that for $d>3$ the analogy with the Van der Waals fluid holds, with the parameters appropriately generalized to depend on the dimension. However we show that for $d=3$ neither the charged nor rotating BTZ black holes exhibit critical behaviour.
While we find the critical exponents in all cases to be the same as for the 4-dimensional case, we uncover interesting new
thermodynamic properties in the Born--Infeld case.  Specifically, 
we  find a term conjugate to the Born--Infeld parameter that we interpret as the Born--Infeld polarization of the vacuum. We correspondingly enlarge the phase space, allowing the Born--Infeld parameter (and its conjugate) to vary. This leads to a complete Smarr relation satisfied by all thermodynamic quantities. 

The outline of our paper is as follows. We first look at higher dimensional Reissner--N\"ordstrom black holes to obtain  the dependence of the various phase space quantities and critical exponents on the spacetime dimensionality $d$. We find that the critical exponents do not depend on the dimension provided $d\geq 4$.  We then consider the Kerr-Newman solution and find in the limit of fixed small angular momenta $J$ that the the same situation holds qualitatively, with the same exact critical exponents.
Turning to the $d=3$ case, we then
 consider charged BTZ black holes. In this case there is no critical phenomena, a situation that persists even if we include rotation.  We then investigate  non-linear electrodynamics via the Born--Infeld action  in four spacetime dimensions. We encounter a rich structure when we investigate the Born--Infeld black hole case. We can have a singularity cloaked by one or two horizons or have a naked singularity depending on the value of the parameters.   We find that in order to satisfy the Smarr relation we need to consider the Born--Infeld parameter as a thermodynamic phase space variable; in so doing, we introduce its conjugate quantity, polarization, and calculate it.   We compute the appropriate Smarr relation and then study the critical behaviour. We close with a concluding section summarizing our results. 
Appendix A reviews the definition of critical exponents and their values for the Van der Waals fluid and the  Reissner--N\"ordstrom-AdS black hole. In appendix B we present an heuristic derivation of the Van der Waals equation in higher dimensions. We gather in appendix C various hypergeometric identities used in the main text.
In what follows we always assume $Q>0$ without loss of generality.

{\bf Note added.} We note here recent work concerned with angular momentum that has some overlap with our paper \cite{Dolan:2012}.


\section{Phase transition of charged AdS black holes in higher dimensions}
\subsection{Charged AdS black holes}

The solution for a spherical charged AdS black hole in $d>3$ spacetime dimensions reads
\ba\label{HDRN}
ds^2 &=& -f dt^2 + \frac{dr^2}{f} + r^{2} d\Omega_{d-2}^2\,,\nonumber\\
F&=&dA\,,\quad A=-\sqrt{\frac{d-2}{2(d-3)}}\frac{q}{r^{d-3}}dt\,.
\ea
Here, $d\Omega_d^2$ stands for the standard element on $S^d$ and function $f$ 
is given by 
\be\label{metfunction}
f = 1 - \frac{m}{r^{d-3}} + \frac{q^2}{r^{2(d-3)}} + \frac{r^2}{l^2}\,.
\ee
Parameters $m$ and $q$ are related to the ADM mass $M$ (in our set up associated with the enthalpy of the system as we shall see) and the black hole charge $Q$ as \cite{ChamblinEtal:1999a}
\ba
M&=&\frac{d-2}{16 \pi} \omega_{d-2} m\,,\nonumber\\
Q&=&\frac{\sqrt{2(d-2)(d-3)}}{8\pi}\,\omega_{d-2}\,q\,,
\ea 
with $\omega_d$ being the volume of the unit $d$-sphere, 
\be
\omega_d=\frac{2\pi^{\frac{d+1}{2}}}{\Gamma\left(\frac{d+1}{2}\right)}\,.
\ee
The metric and the gauge field \eqref{HDRN} solve the Einstein--Maxwell equations following from the bulk action 
\be
I_{EM}=-\frac{1}{16\pi}\int_M d^dx\sqrt{-g}\Bigl(R-F^2+\frac{(d-1)(d-2)}{l^2}\Bigr)\,.
\ee 
In our considerations we interpret the cosmological constant $\Lambda = - \frac{(d-1)(d-2)}{2 l^2}$ as a thermodynamic 
pressure $P$,
\be\label{PLambda}
P = - \frac{1}{8 \pi} \Lambda=\frac{(d-1)(d-2)}{16 \pi l^2}\,.  
\ee
The corresponding conjugate quantity, the thermodynamic volume, is given by \cite{CveticEtal:2010} 
\be\label{volrp}
V  = \frac{\omega_{d-2} {r_+}^{d-1}}{d-1}\,.
\ee
The black hole temperature reads 
\be\label{T4}
T = \frac{f'(r_+)}{4 \pi}
= \frac{d-3}{4 \pi r_{+}} \Bigl(1 - \frac{q^2}{r_+^{2(d-3)}} + \frac{d-1}{d-3}\frac{r_+^2}{l^2} \Bigr)\,,
\ee
with $r_+$ being the position of black hole horizon, determined from $f(r_+)=0$.
The black hole entropy $S$ and the electric potential $\Phi$ (measured at infinity with respect
to the horizon) are  
\ba\label{S4} 
S &=& \frac{A_{d-2}}{4}, \quad A_{d-2} = \omega_{d-2} r_+^{d-2}\,,\\
\Phi&=&\sqrt{\frac{d-2}{2(d-3)}}\frac{q}{r_+^{d-3}}\,.
\ea
All these quantities satisfy the following Smarr formula: 
\be
M=\frac{d-2}{d-3}TS+\Phi Q-\frac{2}{d-3} VP\,,
\ee 
as well as the (extended phase-space) 1st law of black hole thermodynamics 
\be
dM=TdS+\Phi dQ+VdP\,.
\ee
These two are related by a dimensional (scaling) argument; see, e.g., \cite{KastorEtal:2009}. 

In the canonical (fixed charge $Q$) ensemble\footnote{For thermodynamics in the grand canonical (fixed potential $\Phi$) ensemble see \cite{PecaLemos:1999}.} a first order phase transition in the $(T,Q)$-plane, reminiscent of  the liquid--gas phase transition of the Van der Waals fluid was noted quite some time ago in the context of a fixed cosmological constant
 \cite{ChamblinEtal:1999a, ChamblinEtal:1999b}. 
The properties of the corresponding 
critical point have been elaborated, for example, in \cite{NiuEtal:2011}.  More recently, however, it has been pointed out  \cite{KubiznakMann:2012} that at least in 4D the corresponding
critical behaviour has a more  natural interpretation in the extended phase space (where the cosmological constant is treated as a thermodynamic pressure), giving the phase transition in the $(P,T)$-plane, making the 
similarity with the Van der Waals fluid complete. In what follows we shall study this situation in the higher-dimensional case.

\subsection{Critical behaviour}\label{sec:II}
\begin{figure}
\begin{center}
\rotatebox{-90}{
\includegraphics[width=0.39\textwidth,height=0.34\textheight]{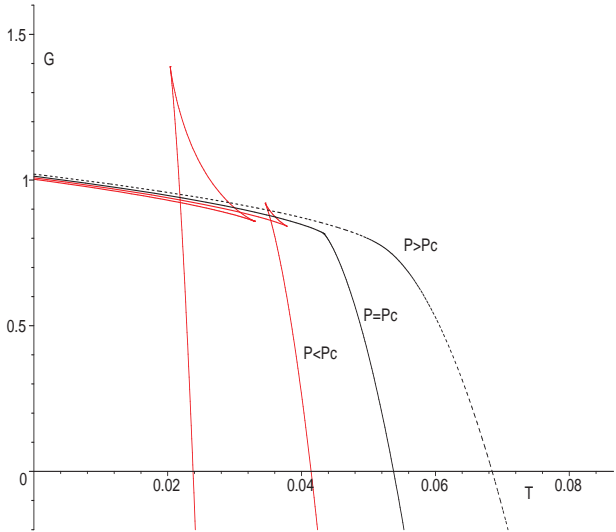}
}
\caption{{\bf Gibbs free energy of charged AdS black hole in $d=4$.}
Gibbs free energy $G$ of 4D RN-AdS black hole is depicted as function of temperature $T$ for fixed $q=1$ and various pressures $P/P_c=1.6, 1, 0.6$ and $0.2$. The behaviour 
for $d>4$ is qualitatively similar, see next figure.
}\label{fig:G4D} 
\end{center}
\end{figure}

\begin{figure}
\begin{center}
\rotatebox{-90}{
\includegraphics[width=0.39\textwidth,height=0.34\textheight]{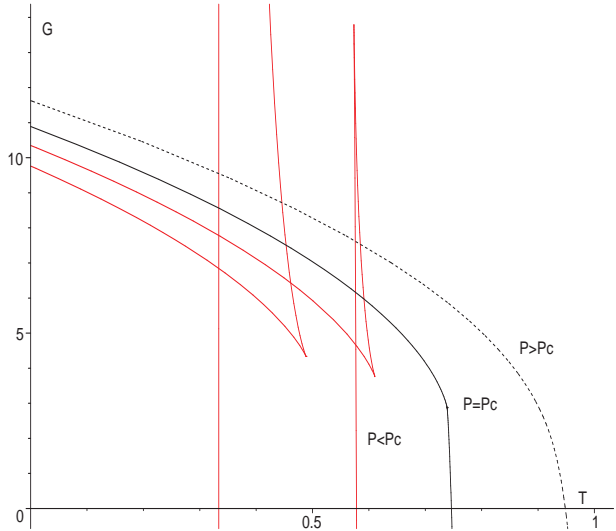}
}
\caption{{\bf Gibbs free energy in $d=10$.}
Gibbs free energy $G$ of 10D RN-AdS black hole is depicted as a function of temperature $T$ for fixed $q=1$ and pressures $P/P_c=1.6, 1, 0.6, 0.2$. 
} \label{fig:G10D} 
\end{center}
\end{figure}

The critical behaviour of a system is captured by its partition function: namely the thermodynamic potential, associated with the 
Euclidean action, calculated at fixed $Q$, fixed $P$ and fixed $T$ is the Gibbs free energy. Using the counterterm method (canceling the AdS divergences) \cite{Mann:1999, EmparanEtal:1999} 
 and  the following boundary term (characteristic for the canonical ensemble \cite{MannRoss:1995}):  
\be
I_b=-\frac{1}{4\pi}\int d^{d-1}x\sqrt{h}F^{ab} n_a A_b\,,
\ee 
the Gibbs free energy $G=M-TS$ reads \cite{ChamblinEtal:1999a} 
\be\label{GibbsQ}
G=G(P,T)
=\! \frac{{\omega}_{d-2} }{16 \pi} \biggl(\!r_+^{d-3} - \frac{16\pi Pr_+^{d-1}}{(d\!-\!1)(d\!-\!2)}
+ \frac{(2d\!-\!5)q^2}{r_+^{d-3}}\biggl).\ \ 
\ee
Here, $r_+$ is understood to be a function of the black hole temperature $T$ and pressure $P$ via equations \eqref{PLambda} and \eqref{T4}; upon  employing  \eqref{volrp} we obtain
\be
P = \frac{T(d-2)}{4 r_{+}} - \frac{(d-3)(d-2)}{ 16 \pi r_+^2}
 + \frac{q^2 (d-3) (d-2)}{ 16 \pi r_{+}^{2(d-2)}}\,.
\ee
The behaviour of $G$ for dimensions $d=4$ and $d=10$ is depicted in Figs.~\ref{fig:G4D} and \ref{fig:G10D}, respectively; we see from these figures 
 the characteristic 1st-order phase transition behaviour.

\begin{figure}
\begin{center}
\rotatebox{-90}{
\includegraphics[width=0.39\textwidth,height=0.34\textheight]{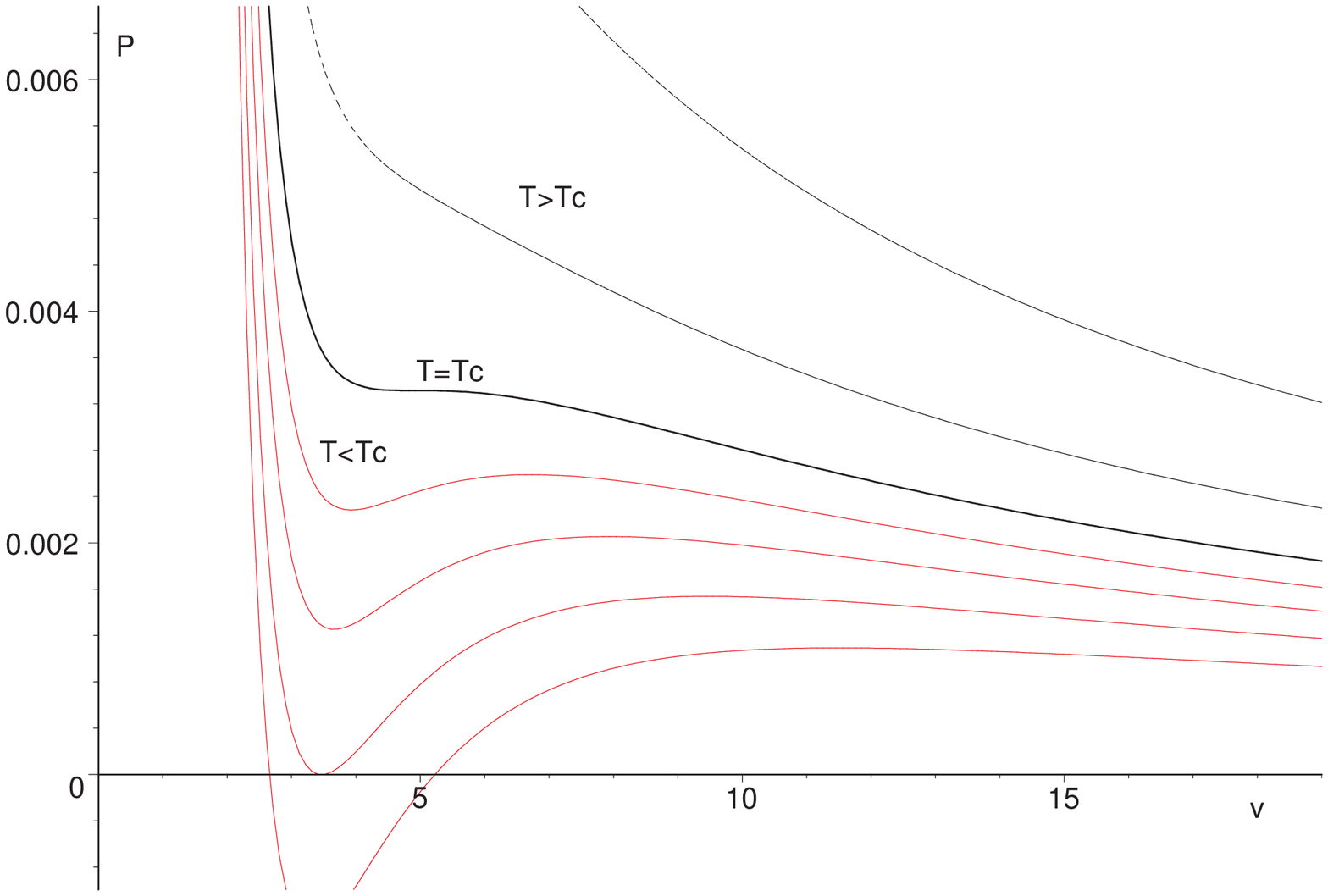}
}
\caption{{\bf $P-v$ diagram of charged AdS black hole in $d=4$.}
The temperature of isotherms decreases from top to bottom. The two upper dashed lines correspond to the ``ideal gas'' one-phase behaviour for $T>T_c$, the critical isotherm $T=T_c$ is denoted by the thick solid line, lower (red) solid lines correspond to two-phase state occurring for $T<T_c$. We have set $q=1$.
The behaviour for $d>4$ is qualitatively similar -- see figure \ref{fig:PV10D}.
}\label{fig:PV4D}  
\end{center}
\end{figure}

\begin{figure}
\begin{center}
\rotatebox{-90}{
\includegraphics[width=0.39\textwidth,height=0.34\textheight]{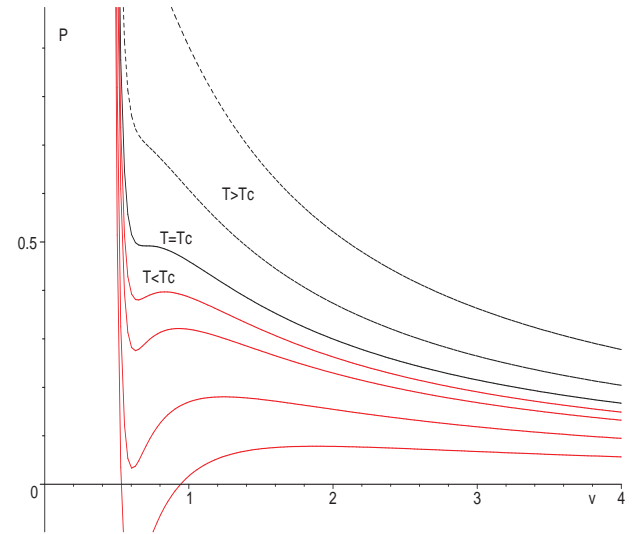}
}
\caption{{\bf $P-v$ diagram in $d=10$.}
The description coincides with that of the previous figure.
}\label{fig:PV10D} 
\end{center}
\end{figure}

To make contact with the Van der Waals fluid in $d$ dimensions (see App.~B), note that  in any dimension $d$ we have $l_P^{d-2} =  G_d\hbar/c^3$, where pressure has units of energy per volume. We thus identify the following relations between our `geometric quantities' $P$ and $T$ and the physical pressure and temperature:
\be
[\mbox{Press}]=\frac{\hbar c}{l_p^{d-2}} [P]\,,\quad [\mbox{Temp}]=\frac{\hbar c}{k} [T]\,.
\ee
Therefore
\ba
\mbox{Press}&=&\frac{\hbar c}{l_p^{d-2}}P 
=\frac{\hbar c}{l_p^{d-2}}\frac{(d-2)T}{4r_+}+\dots\nonumber\\
&=&\frac{k \mbox{Temp}(d-2)}{4 l_p^{d-2} r_+}+\dots
\ea
Comparing with the Van der Waals equation, \eqref{VdWstate2}, we
conclude that we should identify the specific volume $v$ of the fluid with the horizon radius of the 
black hole as 
\be
v = \frac{4 r_+ l_P^{d-2}}{d-2}\,.
\ee
In geometric units we have 
\be
r_+ = \kappa v\,,\quad \kappa=\frac{d-2}{4}\,,  
\ee
and the equation of state reads 
\be\label{statev}
P = \frac{T}{v} - \frac{(d-3)}{\pi (d-2) v^2} + \frac{q^2 (d-3)}{4 \pi v^{2(d-2)} \kappa^{2d-5}}\,.
\ee
The associated $P-v$ diagram\footnote{%
Throughout our paper we display $P-v$ diagrams, with $v$ being the specific volume of the corresponding fluid, rather than $P-V$ diagrams, with $V$ being the thermodynamic volume.  In $d$ dimensions the two are  related via 
\be
V=\frac{\omega_{d-2}}{d-1}\left(\frac{d-2}{4}\right)^{d-1}\!\!v^{d-1}\,.\nonumber
\ee
}
for a 4D black hole is displayed in Fig.~\ref{fig:PV4D}. Obviously, for $T<T_c$ there is a small--large black hole phase transition in the system. This qualitative behaviour persists in higher dimensions---see Fig.~\ref{fig:PV10D} for an illustration in $d=10$. 

Critical points occur at stationary points of inflection in the $P-v$ diagram, where 
\be\label{ddP}
\frac{\partial P}{\partial v} = 0\,,\quad \frac{{\partial}^2 P}{{\partial} v^2} = 0\,,
\ee
yielding
\ba
v_c &=& \frac{1}{\kappa}\left[q^2 (d-2)(2d-5) \right]^{1/[2(d-3)]}\,, \nonumber\\
T_c &=& \frac{(d-3)^2}{ \pi \kappa v_c (2d-5)}\,,\nonumber\\
P_c &=& \frac{(d-3)^2}{16 \pi {\kappa}^2 {v_c}^2}\,.
\ea
This gives the following universal (independent of $q$) ratio
\be\label{PcvcTc}
{\rho}_c=\frac{P_c v_c}{T_c} = \frac{2d-5}{4d-8}\,.
\ee
Note that only in $d=4$ do we recover the ratio $\rho_c=3/8$, characteristic for a Van der Waals fluid in any number of spacetime dimensions.

\begin{figure}
\begin{center}
\rotatebox{-90}{
\includegraphics[width=0.39\textwidth,height=0.34\textheight]{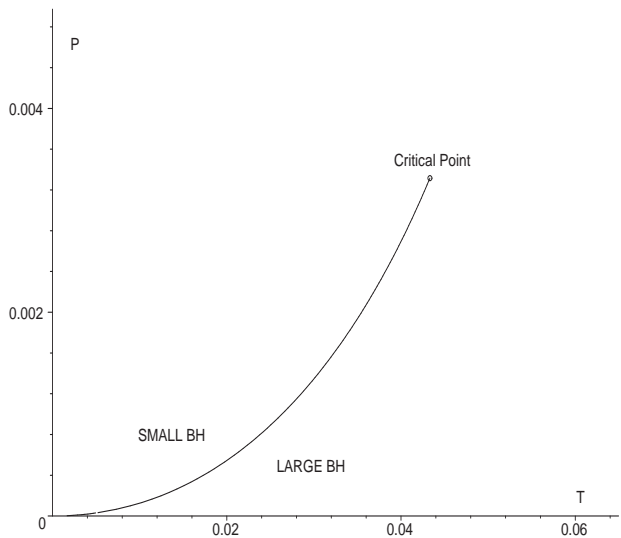}
}
\caption{{\bf Coexistence line of charged AdS black hole in $d=4$.}
This figure displays the coexistence line of the small-large black hole
phase transition of a charged AdS black hole for $d=4$ system in the
$(P, T)$-plane. The critical point is highlighted by a small circle
at the end of the coexistence line. The behaviour for $d>4$ is qualitatively similar, as shown in figure \ref{fig:PT6D}.
}\label{fig:PT4D}  
\end{center}
\end{figure}

\begin{figure}
\begin{center}
\rotatebox{-90}{
\includegraphics[width=0.39\textwidth,height=0.34\textheight]{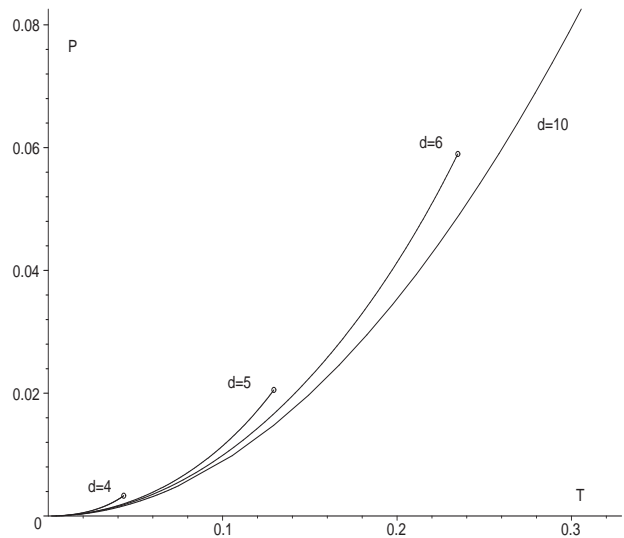}
}
\caption{{\bf Coexistence line in higher dimensions.}
The coexistence line is displayed for $d=4,5, 6$ and $10$. 
}\label{fig:PT6D}  
\end{center}
\end{figure}

The coexistence line in the $(P, T)$-plane is displayed in Figs.~\ref{fig:PT4D}-\ref{fig:PT6D}.

\subsection{Critical exponents}
Following the approach taken previously for $d=4$ \cite{KubiznakMann:2012},
we shall now calculate the critical exponents characterizing the behaviour of physical quantities in the vicinity of the critical point found in the previous section.
The definition of these exponents is for convenience gathered in appendix A.  

To calculate the critical exponent $\alpha$ we consider the entropy $S$, \eqref{S4}, as a function of $T$ and $V$. Using \eqref{volrp} we have 
\be
S=S(T,V)=\omega_{d-2}^{\frac{1}{d-1}}\Bigl[(d-1)V\Bigr]^{\frac{d-2}{d-1}}\,.
\ee 
Since this is independent of $T$, we have $C_V=0$ and hence $\alpha=0$.
Defining 
\be\label{reduced}
p=\frac{P}{P_c}\,,\quad \nu=\frac{v}{v_c}\,,\quad \tau=\frac{T}{T_c}\,,
\ee
the equation of state \eqref{statev} translates into the following `law of corresponding states':
\be\label{statesd}
p=\frac{4(d-2)\tau}{(2d-5)\nu}-\frac{d-2}{(d-3)\nu^2}+\frac{1}{(d-3)(2d-5)\nu^{2d-4}}\,.
\ee
Although this law has explicit dimensional dependence, we shall show that this does not affect the behaviour of the critical exponents. 

Slightly more generally, let us assume (as will be the case for all the examples studied in this paper) that the law of corresponding states takes the form 
\be\label{general}
p=\frac{1}{\rho_c}\frac{\tau}{\nu}+h(\nu)\,,
\ee
where $\rho_c$ stand for the critical ratio.
Expanding this equation near the critical point
\be\label{omegat}
 \tau=t+1\,,\quad \nu=(\omega+1)^{1/z}\,,
\ee
where $z>0$, and using the fact that from the definition of the critical point we have 
\be
\frac{1}{\rho_c}+h(1)=1\,,\quad 
{\rho_c}h'(1)=1\,,\quad 
\rho_c h''(1)=-2\,,
\ee
and so obtain
\be\label{generalexpansion}
p=1+At-Bt\omega-C\omega^3+ O(t\omega^2, \omega^4)\,,
\ee
where
\be\label{ABC}
A=\frac{1}{\rho_c}\,,\quad B=\frac{1}{z\rho_c}\,,\quad 
C=\frac{1}{z^3}\left(\frac{1}{\rho_c}-\frac{h^{(3)}(1)}{6}\right)\,. 
\ee
Let us further assume that $C>0$ (again this is for all of the examples). 
Differentiating the series for a fixed $t<0$ we get  
\be\label{dPgeneral}
dP=-P_c(Bt+3C\omega^2)d\omega\,.
\ee
Employing Maxwell's equal area law, see, e.g.,  \cite{KubiznakMann:2012}, while
denoting $\omega_s$ and $\omega_l$ the `volume' of small and large black
holes, we get the following two equations:
\ba
p&=&1+At-Bt\omega_l-C\omega_l^3=1+At-Bt\omega_s-C\omega^3_s\,,\nonumber\\
0&=&\int_{\omega_l}^{\omega_s} \omega dP\,.
\ea
The unique non-trivial solution is
\be
\omega_s = - \omega_l = \sqrt{\frac{-Bt}{C}}\,,
\ee
yielding
\be
\eta = V_c(\omega_l - \omega_s) = 2 V_c \omega_l\propto \sqrt{-t}  \quad \Rightarrow \quad \beta = \frac{1}{2}\,.
\ee
To calculate the exponent $\gamma$, we use again \eqref{dPgeneral},
to get 
\be
\kappa_T = - \frac{1}{V} \frac{\partial V}{\partial P} \Big |_{T}\propto 
\frac{1}{P_c}\frac{1}{Bt} \quad \Rightarrow \quad \gamma = 1\,.
\ee
Finally, the `shape of the critical isotherm' $t = 0$ is
given by \eqref{generalexpansion}, i.e., 
\be
p - 1 = -C\omega^3 \quad \Rightarrow \quad \delta = 3\,.
\ee
To conclude, provided that the equation of state takes the form \eqref{general}, such that the coefficient $C$, given by \eqref{ABC} is non-trivial and 
positive, we recover the (mean field theory) critical exponents
\be
\beta=\frac{1}{2}\,,\quad \gamma=1\,,\quad \delta=3\,.
\ee

In particular, for charged AdS black holes in any dimension $d$, the law of corresponding states, \eqref{statesd}, takes the form \eqref{general}. Taking $z=d-1$ (in which case $\omega=\frac{V}{V_c}-1$), we obtain the expansion \eqref{generalexpansion} with    
\be\label{ABCd}
A=\frac{4d-8}{2d-5}\,,\quad 
B=\frac{4d-8}{(2d-5)(d-1)}\,,\quad 
C=\frac{2d-4}{3(d-1)^3}
\ee
and so the discussion above applies. We conclude that the thermodynamic exponents associated with the charged AdS black hole in any dimension $d>3$  coincide with those of the Van der Waals fluid.   We now consider the effects of rotation on these results in the following section.

\section{Rotating black holes}
\subsection{Thermodynamics}

The charged AdS rotating black hole solution is given by the Kerr-Newman-AdS metric 
\ba
ds^2&=&-\frac{\Delta}{\rho^2}\left[dt-\frac{a\sin^2\!\theta}{\Xi}d\varphi\right]^2
+\frac{\rho^2}{\Delta} dr^2+\frac{\rho^2}{S}d\theta^2\nonumber\\
&&+\frac{S\sin^2\!\theta}{\rho^2}\left[a dt-\frac{r^2+a^2}{\Xi}d\varphi\right]^2\,,
\ea
in $d=4$, where
\ba
\rho^2&=&r^2+a^2\cos^2\!\theta\,,\quad \Xi=1-\frac{a^2}{l^2}\,,
\quad S=1-\frac{a^2}{l^2}\cos^2\!\theta\,,\nonumber\\
\Delta&=&(r^2+a^2)\left(1+\frac{r^2}{l^2}\right)-2mr+q^2\,.
\ea
The $U(1)$ potential reads
\be
A=-\frac{qr}{\rho^2}\left(dt-\frac{a\sin^2\!\theta}{\Xi}d\varphi\right)\,.
\ee
The thermodynamics of charged rotating black holes treating the cosmological constant as independent variable was performed in 
\cite{CaldarelliEtal:2000, CveticEtal:2010, Dolan:2011a}. In particular, one has the following thermodynamic quantities:
\ba
S&=&\frac{\pi(r_+^2+a^2)}{\Xi}\,,\quad T=\frac{r_+\left(1+\frac{a^2}{l^2}+3\frac{r_+^2}{l^2}-\frac{a^2+q^2}{r_+^2}\right)}{4\pi (r_+^2+a^2)}\,,\nonumber\\
\Phi&=&\frac{qr_+}{r_+^2+a^2}\,,\quad \Omega_H=\frac{a\Xi}{r_+^2+a^2}\,.
\ea
The mass $M$, the charge $Q$, and the angular momentum $J$ are related to parameters $m$, $q$, and $a$ as follows:
\be\label{physical}
M=\frac{m}{\Xi^2}\,,\quad Q=\frac{q}{\Xi}\,,\quad J=\frac{am}{\Xi^2}\,.
\ee
Using the counterterm method \cite{Mann:1999, EmparanEtal:1999}, the action for the canonical ensemble was calculated in \cite{HawkingEtal:1999, CaldarelliEtal:2000} and reads
\be
I=\frac{\beta}{4}\left(r_+-\frac{r_+^3}{\Xi l^2}+\frac{a^2+q^2}{r_+\Xi}+\frac{2q^2r_+}{\Xi(r_+^2+a^2)}\right)\,.
\ee
The corresponding Gibbs free energy reads
\be
G=G(P,T, J, Q)=\frac{I}{\beta}+\Omega J\,, 
\ee
where \cite{CaldarelliEtal:2000}
\be
\Omega=\Omega_H-\Omega_\infty=\Omega_H+\frac{a}{l^2}\,.
\ee
The thermodynamic volume is  \cite{CveticEtal:2010, Dolan:2011b}  
\be
V=\frac{2\pi}{3}\frac{(r_+^2+a^2)(2r_+^2l^2+a^2l^2-r_+^2a^2)+l^2q^2a^2}{l^2\Xi^2 r_+}\,
\ee 
with the pressure $P=\frac{3}{8\pi}\frac{1}{l^2}$ still given by \eqref{PLambda}.
The equation of state is written as
\be
T=\frac{r_+\left(1+\frac{a^2}{l^2}+3\frac{r_+^2}{l^2}-\frac{a^2+q^2}{r_+^2}\right)}{4\pi (r_+^2+a^2)}\,
\ee
where we  express the parameters $a$ and $q$ in terms of the physical quantities $J$ and $Q$, \eqref{physical}. This gives
\ba
P&=&\frac{T}{2r_+}-\frac{1}{8\pi r_+^2}+\frac{Q^2}{8\pi r_+^4}\nonumber\\
&&+\frac{3(4r_+^4+8\pi T r_+^5-8\pi T r_+^3 Q^2+r_+^2Q^2-2Q^4)}{8 \pi r_+^6(r_+^2+2\pi T r_+^3+2Q^2)^2}J^2\nonumber\\
&&+O(J^4)\,
\ea
as an expansion in powers of the angular momentum $J$.
In what follows we expand all quantities to $O(J^2)$, neglecting all terms higher order in $J$.
We further introduce the following quantity (which we shall see corresponds to the specific volume of the associated Van der Waals fluid)
\be
v=2\left(\frac{3V}{4\pi}\right)^{1/3}\!\!\!\!
=2r_++\frac{12(8 \pi r_+^4P+3r_+^2+Q^2)}{r_+(3r_+^2+8\pi r_+^4P+3Q^2)^2} J^2\,,
\ee
and finally obtain  the equation of state 
\ba\label{Pstaterot}
P&=&\frac{T}{v}-\frac{1}{2\pi v^2}+\frac{2Q^2}{\pi v^4}+\frac{48 J^2}{\pi v^6}\nonumber\\
&&-\frac{96Q^2(24Q^2+5v^2+6\pi T v^3)J^2}{\pi v^6(8Q^2+v^2+\pi T v^3)^2}\,.\quad 
\ea

\begin{figure}
\begin{center}
\rotatebox{-90}{
\includegraphics[width=0.39\textwidth,height=0.34\textheight]{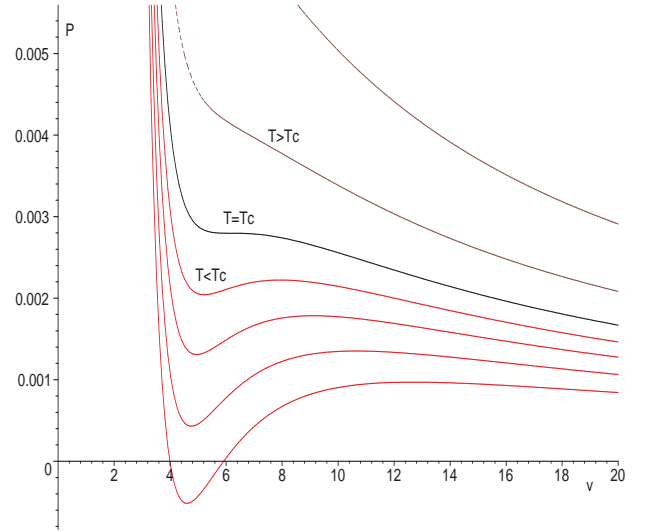}
}
\caption{{\bf $P-v$ diagram for the rotating black hole.}
The isotherms in $P-v$ plane are depicted for a rotating black hole with $J=1$.
}\label{PVKerr}
\end{center}
\end{figure}

\begin{figure}
\begin{center}
\rotatebox{-90}{
\includegraphics[width=0.39\textwidth,height=0.34\textheight]{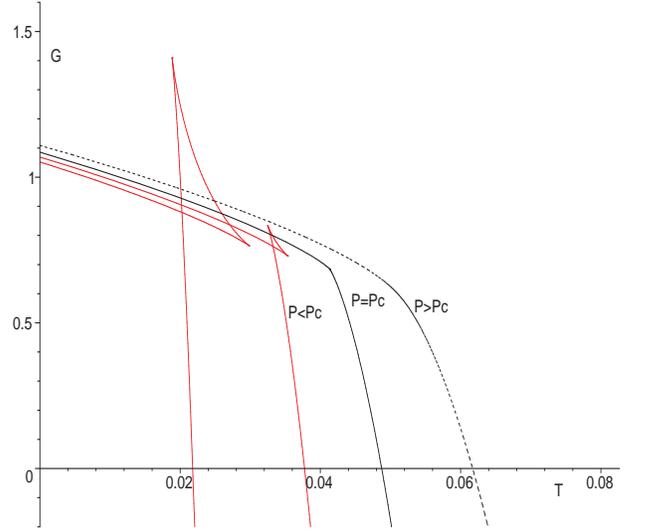}
}
\caption{{\bf Gibbs free energy for the rotating black hole.}
Isobars of the Gibbs free energy for a rotating black hole with $J=1$ are depicted for 
$P/P_c=1.6, 1, 0.6, 0.2$.
} \label{GKerr}
\end{center}
\end{figure}

\begin{figure}
\begin{center}
\rotatebox{-90}{
\includegraphics[width=0.39\textwidth,height=0.34\textheight]{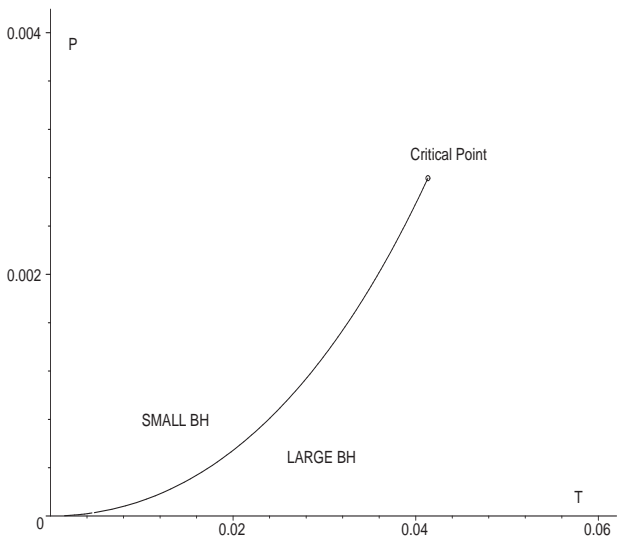}
}
\caption{{\bf Coexistence line for the rotating black hole.}
The coexistence line of the phase transition between small and large slowly rotating black holes. The figure is qualitatively similar to charged AdS black hole case.
}  \label{PTKerr}
\end{center}
\end{figure}

In what follows we limit ourselves to the case of zero charge $Q=0$, for which \eqref{Pstaterot} becomes
\cite{Dolan:2011b}
\be\label{PstateQ0}
P=\frac{T}{v}-\frac{1}{2\pi v^2}+\frac{48 J^2}{\pi v^6}\,.
\ee
In the same approximation we get the following expression for the Gibbs potential:
\be\label{GrotQ0}
G=\frac{v}{8}-\frac{\pi P v^4}{12}+\frac{20 J^2}{v^3}\,.
\ee
Upon comparing \eqref{PstateQ0} to \eqref{statev} and  \eqref{GrotQ0} to \eqref{GibbsQ} we note the same
qualitative behaviour as in the charged case, with fixed $J$ playing the role of fixed $Q$.  However the quantitative
dependence of $P$ on $v$ is different.  We depict the
 $P-v$ diagram in Fig.~\ref{PVKerr} and  the Gibbs potential in Fig.~\ref{GKerr}, where the qualitative behaviour is seen to
 be similar to the non-rotating charged case.  The  $P-T$ diagram is illustrated in Fig.~\ref{PTKerr}.

\subsection{Critical behaviour}
 Using \eqref{ddP} the critical point occurs at 
\ba
v_c&=&2\times 90^{1/4}\sqrt{J}\,,\nonumber\\
T_c&=&\frac{90^{3/4}}{225\pi} \frac{1}{\sqrt{J}}\,,\nonumber\\
P_c&=&\frac{1}{12\sqrt{90} \pi}\frac{1}{J} 
\ea
and these relations in turn yield the universal critical ratio
\be
\rho_c=\frac{P_c v_c}{T_c}=\frac{5}{12}\,.
\ee 
that differs from the Van der Waals fluid. Upon inclusion of higher powers of $J$ we expect this to become $J$ dependent. We shall see a similar phenomenon take place in the Born--Infeld case. 

Defining the quantities $p, \nu$ and $\tau$ as previously, \eqref{reduced}, we get the following law of corresponding states:
\be
p=\frac{12}{5} \frac{\tau}{\nu}-\frac{3}{2} \frac{1}{\nu^2}+\frac{1}{10} \frac{1}{\nu^6}\,,
\ee 
which is again universal and of the form \eqref{general}. Expanding around the critical point using the variables $\tau$ and $\omega$ with $z=3$, we get
\be
p=1+\frac{12}{5}t-\frac{4}{5}t \omega -\frac{2}{27}\omega^3 +O(t \omega^2, \omega^4)\,.
\ee
It is now obvious from this expansion  and the discussion in Sec.~II.C that the critical exponents $\beta, \delta$ and $\gamma$
do not change, i.e., we have
\be
\beta=\frac{1}{2}\,,\quad \gamma=1\,, \quad
\delta=3\,.
\ee 
Moreover, we have
\be
S=S(T,V)=\frac{\pi(r_+^2+a^2)}{\Xi}=\frac{\pi v^2}{4}+O(J^4)\,.
\ee
Hence to  quadratic order in $J$ we have 
\be
C_v= T \left(\frac{\partial S}{\partial T}\right)_v=0\,,
\ee
which implies that the critical exponent $\alpha=0$. So we conclude that (at least for slowly rotating black holes)
the critical point belongs to the same universality class as the one of the Van der Waals fluid.
 
One might wonder if this continues to hold also for $d=3$. We show  in the next section that this is not the case, and  the BTZ black hole is rather different from its charged/rotating counterparts in higher dimensions.

\section{BTZ black holes}
In this section we investigate the case of a charged BTZ black hole. This case needs to be considered separately due to the behaviour of the electromagnetic field in three spacetime dimensions.
The metric and the gauge field are given by \cite{BanadosEtal:1992}
\ba
ds^2&=&-fdt^2+\frac{dr^2}{f}+r^2d\varphi^2\,,\nonumber\\
F&=&dA\,,\quad A=-Q\log\left(\frac{r}{l}\right)dt\,
\ea
with
\be
f = -M- \frac{Q^2}{2} \log \left( \frac{r}{l} \right)+\frac{r^2}{l^2}\,.  
\ee
With this choice we have a solution of the Einstein--Maxwell system with the cosmological constant $\Lambda=-1/l^2=-8\pi P$.
The temperature is
\ba\label{BTTemp}
T = \frac{f^{'} (r_+)}{4 \pi} = \frac{r_+}{2\pi l^2} - \frac{Q^2}{8\pi r_+}\,,
\ea
which translates to the following equation of state:
\be
P = \frac{T}{v} + \frac{Q^2}{2\pi v^2}\,,\quad v=4r_+\,.
\ee
It is obvious that such an equation does not admit any inflection point and hence the charged BTZ black hole does not exhibit any critical behaviour.
This is quite analogous to the  toroidal (planar) case observed in four dimensions \cite{KubiznakMann:2012}.

One might wonder whether this remains true when one includes rotation. The corresponding metric was   
constructed in \cite{MartinezEtal:2000} and  is rather complicated. Let us limit ourselves to considering the uncharged rotating 
BTZ black hole. 
The rotating BTZ black hole is given by \cite{BanadosEtal:1992}
\ba
ds^2&=&-fdt^2+\frac{dr^2}{f}+r^2\left( -\frac{J}{2r^2}dt+d\varphi\right)^2\,,\nonumber\\
f&=&-M+\frac{r^2}{l^2}+\frac{J^2}{4r^2}\,.
\ea
The corresponding equation of state reads
\be
P=\frac{T}{v}+\frac{8J^2}{\pi v^4}\,,\quad v=4 r_+\,.
\ee
Similar to the charged case it does not admit  critical behaviour. We expect this to remain true also in the general rotating case with nontrivial charge 
\cite{MartinezEtal:2000}. 

To conclude, for BTZ black holes, contrary to the higher-dimensional case, we do not observe any critical behaviour.

\section{Influence of nonlinear electrodynamics} 

\subsection{Born--Infeld-AdS solution} 

The Einstein--Born--Infeld theory in AdS is described by the following bulk action \cite{BornInfeld:1934}  
\ba\label{IEM}
I_{EM}&=&-\frac{1}{16\pi}\int_M \sqrt{-g}\left(R+{\cal L}_{BI}+\frac{6}{l^2}\right)\,,\nonumber\\
{\cal L}_{BI}&=&4{b^2}\Bigl(1-\sqrt{1+\frac{2F}{b^2}}\ \Bigr)\,,\quad F=\frac{1}{4}F^{ab}F_{ab}\,, \quad
\ea
where we restrict ourselves to $d=4$.
The parameter $b$ represents the maximal electromagnetic field strength. With motivations from string theory, e.g. \cite{Gibbons:2001}, $b$ can also be related to string tension, $b=\frac{1}{2\pi\alpha'}$.

In Schwarzschild-like coordinates the metric and the electromagnetic field of the spherically symmetric solution read  
\cite{FernandoKrug:2003, Dey:2004, CaiEtal:2004} 
\ba
ds^2&=&-fdt^2+\frac{dr^2}{f}+r^2d\Omega_2^2\,,\label{BImetric}\\
F&=&Edt\wedge dr\label{BIE}\,, \quad E=\frac{Q}{\sqrt{r^4+Q^2/b^2}}\,.
\ea 
Here, $d\Omega^2_2$ stands for the standard element on $S^2$ and the function $f$ is given by
\ba
f&=&1-\frac{2M}{r}+\frac{r^2}{l^2}+\frac{2b^2}{r}\int_r^\infty \Bigl(\sqrt{r^4+\frac{Q^2}{b^2}}-r^2\Bigr)dr\nonumber\\
&=&1-\frac{2M}{r}+\frac{r^2}{l^2}+\frac{2b^2r^2}{3}\Bigl(1-\sqrt{1+\frac{Q^2}{b^2r^4}}\Bigr)\nonumber\\
&&+\frac{4Q^2}{3r^2}\,{}_2F_1\left(\frac{1}{4},\frac{1}{2}; \frac{5}{4};-\frac{Q^2}{b^2r^4}\right)\,,
\ea
with ${}_2F_1$ being the hypergeometric function.  
Note that whereas the hypergeometric series expression is convergent only for $|z|<1$, i.e.,  for $r>r_0=\sqrt{Q/b}$, one may use the integral representation, \eqref{C6} (discussed further in the appendix)  for any $r>0$.

\begin{figure}
\begin{center}
\rotatebox{-90}{
\includegraphics[width=0.39\textwidth,height=0.34\textheight]{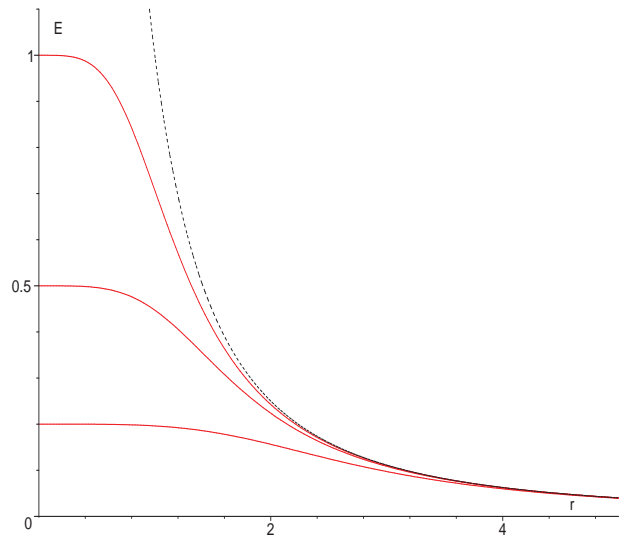}}
\caption{{\bf BI electric field strength.}
The field strength $E$ is depicted as a function of $r$ for various values $b$ and fixed $Q=1$. The value of $b$ decreases from top to bottom. The lower three red solid lines correspond to $b=0.2$, $b=1/2$, and $b=1$, respectively. The field reaches the finite value in the origin in these cases. 
The top (black) dashed line corresponds to the limiting Maxwell case ($b\to \infty$). 
}  \label{fig:E}
\end{center}
\end{figure}

The parameter $M$ represents the ADM mass and parameter $Q$ the asymptotic charge of the solution. 
The electric field strength is depicted in Fig.~\ref{fig:E}.
The structure of the horizons is discussed in the next subsection.

\subsection{Two types of black hole solutions}

The solution \eqref{BImetric} possesses a singularity at $r=0$. This singularity is cloaked by one, or two horizons, or describes a naked singularity, depending on the value of the parameters. 
There are two types of black hole solutions as sketched in Fig.~\ref{fig:BHs}.  This can be seen from the expansion of $f$ around $r=0$. 
Using the formula \eqref{expansionzero} derived in App.~C, we find \cite{FernandoKrug:2003, Fernando:2006}
\be
f=1-\frac{2(M-M_m)}{r}-{2}bQ+O(r)\,,
\ee
where 
\be
 M_m=\frac{1}{6}\sqrt{\frac{b}{\pi}}Q^{3/2}\Gamma\Bigl(\frac{1}{4}\Bigr)^2\, 
\ee
is the {\em `marginal mass'}. 
Depending on the value of $M$, $b$, $l$, and $Q$ we have the following cases:
\begin{figure}
\begin{center}
\rotatebox{-90}{
\includegraphics[width=0.39\textwidth,height=0.34\textheight]{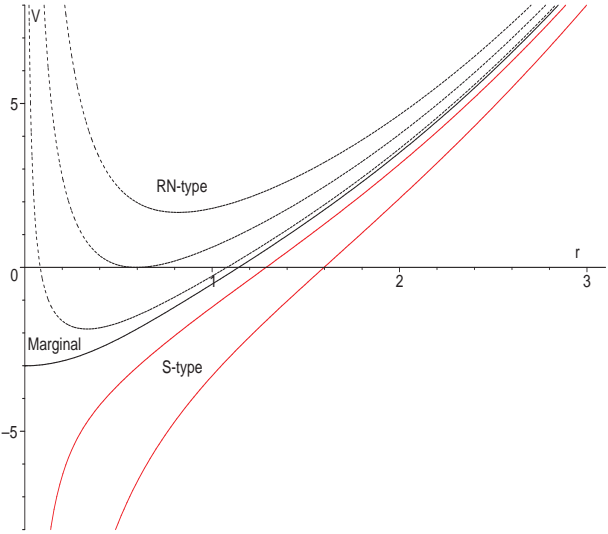}
}\label{Fig4}
\caption{{\bf Types of Born--Infeld-AdS black holes.}
Two types of possible BI-AdS black hole solutions are displayed for $b=2$. The marginal case $M=M_m$ is highlighted by a thick solid line. The (red) solid lines below it
represent the S-type of $M>M_m$. The upper dashed lines correspond to various cases of RN-type. The naked singularity, extremal BH, and two horizon solution of this type are displayed from top to bottom. The parameters were chosen $Q=1$ and  $l=1$. 
} \label{fig:BHs}
\end{center}
\end{figure} 
\begin{figure}
\begin{center}
\rotatebox{-90}{
\includegraphics[width=0.39\textwidth,height=0.34\textheight]{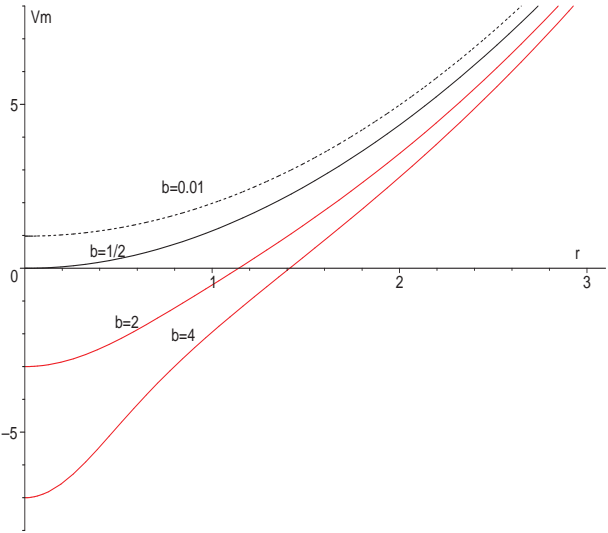}
}\label{Fig5}
\caption{{\bf Marginal case.}
The sequence of marginal metric functions $f_m$ is depicted for various values of $b$ and fixed $Q=1,\ l=1$. As $b$ decreases the marginal line moves upwards, shifting the $f$ of the RN-type to more positive values. For $bQ<1/2$ the corresponding $f$ is necessarily positive and hence the RN-type describes a naked singularity. 
} \label{fig:Marginal} 
\end{center}
\end{figure}

\begin{figure}
\begin{center}
\rotatebox{-90}{
\includegraphics[width=0.39\textwidth,height=0.34\textheight]{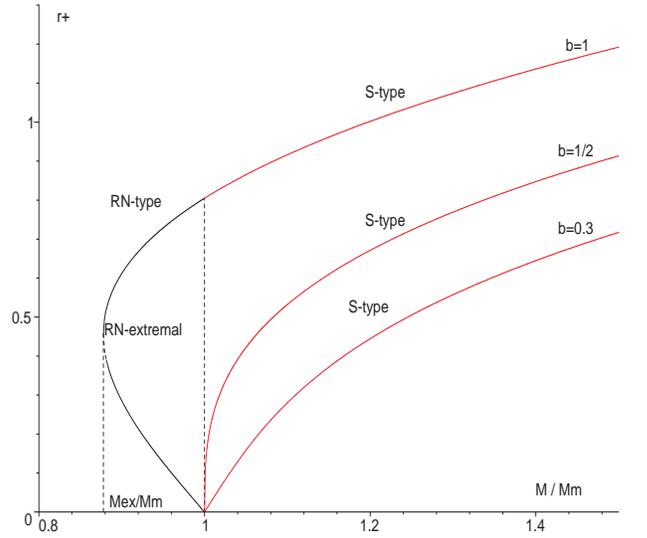}
}
\caption{{\bf Black hole horizon radia.}
The horizon radii are depicted for various types as a function of $M/M_m$. Whereas for $bQ<1/2$ only S-type black hole exists (lower red curve), for $b>1/2$ the S-type  smoothly joins the RN-type (black) and continues till extremal mass $M_{ex}/M_m$. The lower black dotted line corresponds to the position of inner horizon of RN-type. The units were chosen $Q=1\,,\ l=1$.
}\label{fig:Mr}  
\end{center}
\end{figure}

For $M>M_m$ we have the {\em `Schwarzschild-like' (S) type}. This type of black hole is characterized by the existence of a spacelike singularity and possesses one horizon. The characteristic behaviour of $f$ is displayed in Fig.~\ref{fig:BHs} by (red) solid lines. The {\em `Reissner--N\"ordstrom' (RN) type} is characterized by $M<M_m$. Similar to the Reissner--N\"ordstrom solution, this may have zero, one, or two horizons, see Fig.~\ref{fig:BHs} and Fig.~\ref{fig:Marginal}. 
The case $M=M_m$ is the {\em `marginal' case,}  for which function $f_m=f(M_m)$ approaches the (finite) value 
$f_m(0)=1-2bQ$. When this is positive, i.e., for     
$bQ\leq \frac{1}{2}\,,$ 
the RN phase describes a naked singularity and the only possible black hole solution is the S-type for $M>M_m$. On the other hand   
when 
\be
bQ> \frac{1}{2}\,,
\ee
the RN-type describes a black hole for $M_{ex}\leq M<M_m$, with 
$M_{ex}$ being the mass of the extremal RN-type black hole, determined from $V(r=r_{ex})=0$, with $r_{ex}$ given by   
\be
1+\left(2b^2+\frac{3}{l^2}\right)r_{ex}^2-2b\sqrt{r_{ex}^4b^2+Q^2}=0\,.
\ee
We gather the possibilities for BI-AdS black holes in the following table; 
the corresponding dependence of $r_+$ on $M/M_m$ is depicted in Fig.~\ref{fig:Mr}:
\be\label{A1}
\begin{array}{|c|c|c|}
\hline
\multicolumn{3}{|c|}{\mbox{BI-AdS black holes}} \\
\hline 
\mbox{Condition(s)} & \mbox{Type} & \mbox{\# horizons}\\
\hline
M>M_m & \mbox{S} & \mbox{one}\\
\hline
M_{ex}<M<M_m  & \mbox{RN}  & \mbox{two} \\
bQ>\frac{1}{2}
 &   & \\
\hline
M=M_{ex}<M_m &  \mbox{RN} & \mbox{one}\\
bQ>\frac{1}{2}
 &  \mbox{(extremal)} & \\
\hline	
\end{array} 
\ee
In all other cases we have a naked singularity. 
Let us stress that $M_m$ depends on $b$ and $Q$, whereas $M_{ex}=M_{ex}(b,Q,l)$. Therefore, for example, 
$M_{ex}<M_m$ imposes a nontrivial restriction on the parameters of the solution.
We also emphasize that  the possibility of S-type BI-AdS black holes is sometimes completely ignored in a consideration
of BI-AdS black hole thermodynamics \cite{MyungEtal:2008}.  We shall see that it is the S-type black holes for which the most interesting behaviour occurs.

\subsection{First law, Smarr formula,  and vacuum polarization}
The Smarr relation follows from the first law of black hole thermodynamics 
and a scaling dimensional argument \cite{KastorEtal:2009}. [See also \cite{KastorEtal:2010} for a generalization to the Lovelock gravities.]  Beginning with {\em Euler's theorem} (for simplicity formulated only for two variables): given a function $g(x,y)$ such that $g(\alpha^p x, \alpha^q y)=\alpha^rg(x,y)$, it follows that 
\be
rg(x,y)=px \left(\frac{\partial g}{\partial x}\right)+qy\left(\frac{\partial g}{\partial y}\right)\,.
\ee 
To obtain the proper scaling relations, for the Born--Infeld case we must consider
the mass of the black hole $M$ to be a function of entropy, pressure, angular momentum, charge, and Born--Infeld parameter $b$, $M=M(S,P, J, Q, b)$. Performing the dimensional analysis, we find $[M]=L\,, \ [S]=L^2\,,\ [P]=L^{-2}\,,\ [J]=L^2\,,\ [Q]=L\,,\ [b]=[E]=L^{-1}\,.$ 
Since $b$ is a dimensionful parameter, the corresponding term will inevitably appear in the Smarr formula. It is also natural to include its variation in the first law.
Hence we have 
\ba\label{Mpom}
M&=&2S \left(\frac{\partial M}{\partial S}\right)-2P\left(\frac{\partial M}{\partial P}\right)
+2J\left(\frac{\partial M}{\partial J}\right)\nonumber\\
&&+ Q\left(\frac{\partial M}{\partial Q}\right)-b\left(\frac{\partial M}{\partial b}\right)\,.
\ea  
Defining further a quantity conjugate to $b$,
\be\label{B}
{\cal B}=\left(\frac{\partial M}{\partial b}\right)\,,
\ee
the first law takes the form\footnote{In fact, this form of the first law remains to be proved. In order to do so, one could, for example, employ the Hamiltonian perturbation theory techniques, similar to \cite{KastorEtal:2010}.}   
\be\label{dM}
dM=TdS+VdP+\Omega dJ+\Phi dQ+{\cal B}db\,.
\ee
Using this and Eq. \eqref{Mpom}, we find the following generalized Smarr formula for stationary Born--Infeld-AdS black holes:
\be\label{Smarr}
M=2(TS-VP+\Omega J)+\Phi Q -{\cal B}b\,.
\ee
Let us emphasize that even in the case when the cosmological constant $\Lambda$ and the parameter $b$ are not varied in the first law, i.e., we have $dM=TdS+\Omega dJ+\Phi dQ$, the Smarr relation \eqref{Smarr} is valid and the $VP$ and $b{\cal B}$ terms therein are necessary for it to hold. 
This resolves the problem of inconsistency of the first law and the corresponding Smarr relation raised by 
Rasheed \cite{Rasheed:1997}; see also \cite{Breton:2004, Huan:2010}.

In particular, for our (static) black hole solution \eqref{BImetric}
we have the following thermodynamic quantities:
The black hole temperature and the corresponding entropy are
\ba
T&=&\frac{1}{4\pi r_+}\!\left[1\!+\!\frac{3r_+^2}{l^2}\!+\!2b^2r_+^2\Bigl(1\!-\!\sqrt{1\!+\!\frac{Q^2}{b^2r_+^4}}\Bigr)\right],
\quad \label{T}\\
S&=&\frac{A}{4}\,,\quad A=4\pi r_+^2\,.\label{S}
\ea
The thermodynamic volume $V$ and the corresponding pressure $P$ are  
\be\label{V}
V=\frac{4}{3}\pi r_+^3\,,\quad P=-\frac{1}{8\pi} \Lambda=\frac{3}{8\pi}\frac{1}{l^2}\,.
\ee
The electric potential $\Phi$, measured at infinity with respect to the horizon, is 
\be
\Phi=\int_{r_+}^\infty \frac{Q dx}{\sqrt{x^4+Q^2/b^2}}
=\frac{Q}{r_+}\,{}_2F_1\!\left(\frac{1}{4},\frac{1}{2};\frac{5}{4};-\frac{Q^2}{b^2r_+^4}\right)\,.
\ee 
Using the first law \eqref{dM} and the formula \eqref{C13}, we get the following expression for the quantity ${\cal B}$:
\ba
{\cal B}&=&\frac{2}{3}br_+^3-\frac{2}{3}br_+^3\sqrt{1+\frac{Q^2}{b^2r_+^4}}+\frac{Q^2}{3br_+\sqrt{1+\frac{Q^2}{b^2r_+^4}}}\nonumber\\
&&+\frac{2}{15}\frac{Q^4}{b^3r_+^5}\,{}_2F_1\left(\frac{5}{4},\frac{3}{2}; \frac{9}{4};-\frac{Q^2}{b^2r_+^4}\right)\,.
\ea
Alternatively, one could calculate ${\cal B}$ from the Smarr formula \eqref{Smarr}, getting 
\be
{\cal B}=\frac{2}{3}br_+^3\Bigl(1-\sqrt{1\!+\!\frac{Q^2}{b^2r_+^4}}\Bigr)
+\frac{Q^2}{3br_+}\,{}_2F_1\!\left(\frac{1}{4},\frac{1}{2}; \frac{5}{4};-\frac{Q^2}{b^2r_+^4}\right)\!.
\ee
The two expressions of course coincide and are related by \eqref{A5}.
We shall return to this quantity later in this section.

\subsection{Critical behaviour}
The thermodynamic behaviour of Born--Infeld-AdS black holes and the corresponding phase transitions were studied in the grand canonical ensemble in \cite{Fernando:2006} and in canonical ensemble in \cite{FernandoKrug:2003}. 
The critical exponents were recently calculated in \cite{BanerjeeEtal:2010, LalaRoychowdhury:2011, BanerjeeRoychowdhury:2012, BanerjeeRoychowdhury:2012b}. In this subsection we discuss the thermodynamics in canonical ensemble in the extended phase space. We also correct the values of the critical exponents and some of the previous results appearing in the literature.
In particular, 
we limit ourselves to the case when $b$ is fixed and consider $P-v$ extended phase space. Our aim is to study the influence of the nonlinearity of the electromagnetic field on the existence of the critical point and its corresponding behaviour.

\subsubsection{Equation of state}
\begin{figure}
\begin{center}
\rotatebox{-90}{
\includegraphics[width=0.39\textwidth,height=0.34\textheight]{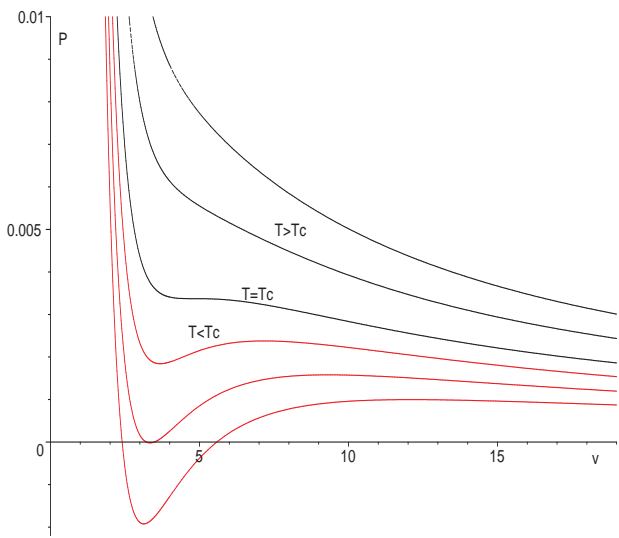}
}
\caption{{\bf $P-v$ diagram of BI-AdS black hole for $b>b_2$.}
The temperature of isotherms decreases from top to bottom. The two upper dashed lines correspond to the ``ideal gas'' one-phase behaviour for $T>T_c$, the critical isotherm $T=T_c$ is denoted by the thick solid line, lower solid lines correspond to temperatures smaller than the critical temperature.
We have set $Q=1$ and $b=1$ for which $T_c\approx 0.04362$.  
The figure is very much reminiscent of the corresponding behaviour of RN-AdS black hole. For $b>b_3\approx 1.72846/Q$ the critical point occurs for the RN-type black hole whereas for $b<b_3$ for the S-type. In both cases the behaviour remains qualitatively the same.
}\label{fig:rPa}  
\end{center}
\end{figure}

\begin{figure}
\begin{center}
\rotatebox{-90}{
\includegraphics[width=0.39\textwidth,height=0.34\textheight]{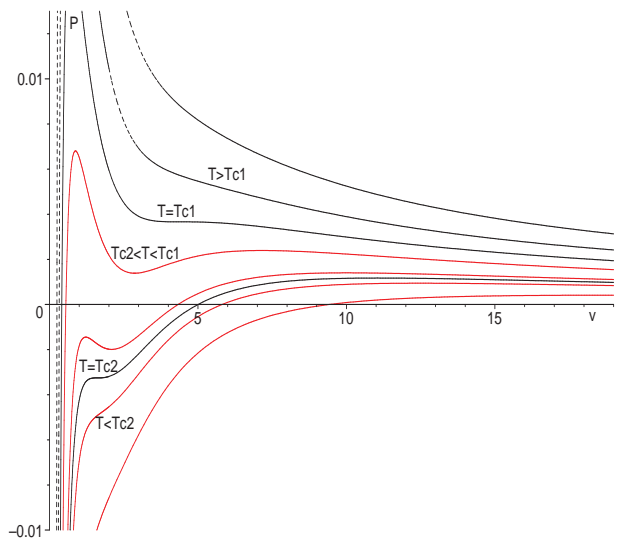}
}
\caption{{\bf $P-v$ diagram of BI-AdS BH for $b\in(b_1, b_2)$.}
We have now two critical points, one at positive pressure, the other at negative pressure. The upper one occurs at bigger horizon radius, bigger mass, and higher temperature. Both are associated with the S-type of black hole solutions.  We have set $Q=1$ and $b=0.45$ for which $T_{c_1}\approx 0.04517$ and $T_{c_2}\approx 0.026885$. As we decrease $b$ further, the critical points start to move towards each other. At $b=b_1\approx 0.4237/Q$ both have positive pressure, see Fig.~\ref{fig:Tc12}. Decreasing $b$ even further, the two critical points eventually merge and disappear. That is for $b<b_0\approx 0.35355/Q$ there are no critical points anymore, Fig.~\ref{fig:Tc12c}. 
} \label{fig:Tc12a} 
\end{center}
\end{figure}

\begin{figure}
\begin{center}
\rotatebox{-90}{
\includegraphics[width=0.39\textwidth,height=0.34\textheight]{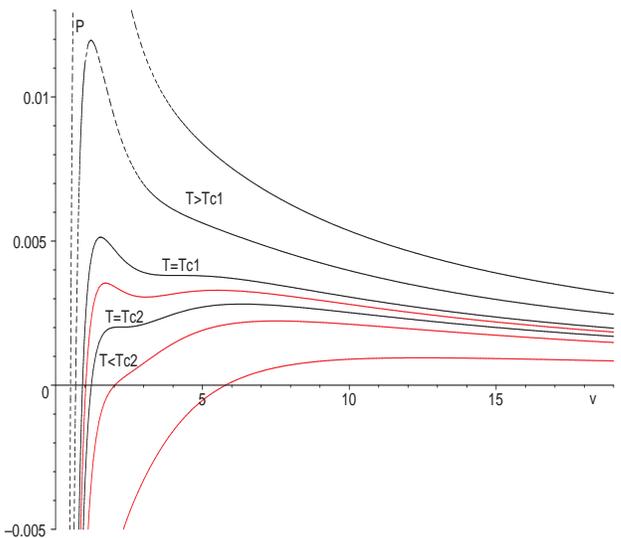}
}
\caption{{\bf $P-v$ diagram of BI-AdS BH for $b\in(b_0, b_1)$.}
We have now two critical points at positive pressure. The upper one has higher radius, temperature, and mass. 
We have set $Q=1$ and $b=0.4$ for which $T_{c_1}\approx 0.045890$ and $T_{c_2}\approx 0.040492$. 
}\label{fig:Tc12}   
\end{center}
\end{figure}

\begin{figure}
\begin{center}
\rotatebox{-90}{
\includegraphics[width=0.39\textwidth,height=0.34\textheight]{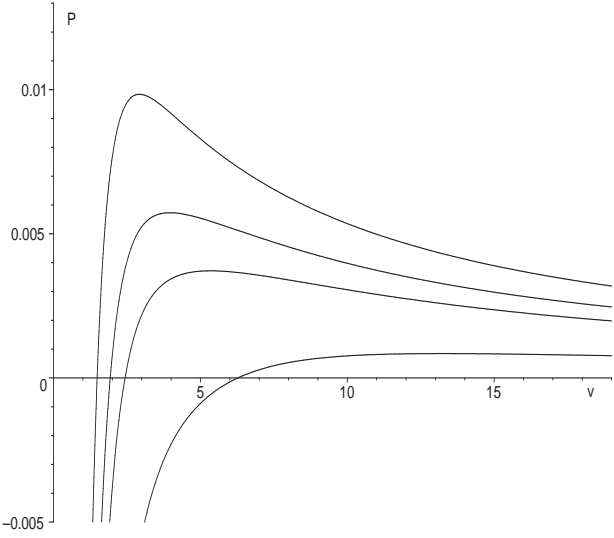}
}
\caption{{\bf $P-v$ diagram of BI-AdS BH for $b<b_0$.}
There are no longer any critical points below $b_0\approx 0.35355/Q$.  We have set $Q=1$ and $b=0.2$.
} \label{fig:Tc12c} 
\end{center}
\end{figure}

For a fixed charge $Q$, Eq. \eqref{T} translates into the 
{\em equation of state} for a Born--Infeld AdS black hole,  $P=P(V,T)$,
\be\label{BIstate}
P=\frac{T}{2r_+}-\frac{1}{8\pi r_+^2}-\frac{b^2}{4\pi}\Bigl(1-\sqrt{1+\frac{Q^2}{b^2r_+^4}}\Bigr)\,.\quad 
\ee
Similar to the Maxwell 4D case, discussed in Sec.~II, it is the horizon diameter $2r_+$, rather than the thermodynamic volume $V$, that should be associated with the specific volume of the corresponding fluid; $v\equiv 2r_+ l_P^2$.  Taking $l_P=1$, Eq. \eqref{BIstate} gives 
\begin{equation}\label{BIstateV}
P =\frac{T}{v}-\frac {1}{2\pi\,{v}^{2}}-\frac{{b}^{2}}{4{\pi}} \Bigl( 1- \sqrt{1+\frac{16 {Q}^{2}}{{b}^{2}{v}^{4}}} \,\Bigr)\,.
\end{equation}
The corresponding $P-v$ diagrams are displayed in Figs.~\ref{fig:rPa}--\ref{fig:Tc12c} for various values of $b$. 

The behaviour of the isotherms depends crucially on how `deep' we are in the Born--Infeld regime. For $bQ>1/2$ we have the `{Maxwellian regime}' and the behaviour is qualitatively similar to  that of a Reisner-Nordstrom-AdS black hole. Namely, for $Q\neq 0$  and for $T<T_c$ there is an inflection point and the behaviour is reminiscent  of the Van der Waals gas, as shown in Fig.~\ref{fig:rPa}. 

However, for $bQ<1/2$ we have much `stranger behaviour', with one, two, or zero inflection points, as shown in Figs.~\ref{fig:Tc12a}-\ref{fig:Tc12c}.  
The {\em critical point} is obtained from Eq.~\eqref{ddP}, 
which leads to the cubic equation 
\be\label{cubic}
x^3 + px +q = 0 \,,
\ee
where
\be
p=-\frac{3 b^2}{32 Q^2}\,,\ q= \frac{b^2}{256 Q^4}\,,\ 
x = \left(v_c^4 + \frac{16 Q^2}{b^2}\right)^{-1/2}\!\!\!\!.
\ee
In order that $v_c=\Bigl( \frac{1}{x^2} - \frac{16 Q^2}{b^2} \Bigr)^{\frac{1}{4}}$ be positive, 
we require an additional constraint 
\be\label{additional}
|x|\leq \frac{b}{4Q}\,.
\ee

To write an analytic expression for $v_c$ in terms of  trigonometric and/or hyperbolic functions let us analyze the cubic. 
The cubic of the form \eqref{cubic}, with $p<0$ and real $q$, admits three or one real roots. Three real roots occur when $ 4 p^3 + 27 q^2 \leq 0 $, i.e., when 
\be
b\geq b_0=\frac{1}{\sqrt{8}Q}\approx\frac{0.35355}{Q}\,,
\ee
in which case the roots are given by 
\begin{eqnarray}\label{xk}
x_k &=& 2\sqrt{-\frac{p}{3}}\cos\left(\frac{1}{3}\arccos\left(\frac{3q}{2p}\sqrt{\frac{-3}{p}}\right)-\frac{2\pi k}{3}\right)\,, \nonumber \\ 
&&\qquad \qquad \qquad \qquad \qquad \qquad k=0, 1, 2\,. 
\label{eq:sol1}
\end{eqnarray} 
It is easy to verify that the additional condition \eqref{additional} is never satisfied for $x_2$, 
always satisfied for $x_1$, and satisfied for $x_0$ provided that 
\be
\frac{1}{\sqrt{8}Q} = b_0\leq b\leq b_2=\frac{1}{2Q}\,.
\ee
One real root occurs for $b<b_0$, in which case 
the solution is given by
\be
x_3 = -2\sqrt{-\frac{p}{3}}\cosh \left(\frac{1}{3} \operatorname{arccosh} \left(-\frac{3q}{2p}\sqrt{\frac{-3}{p}}\right)\right)\,, 
\ee 
which, however, does not satisfy \eqref{additional}\,.

\begin{figure}
\begin{center}
\rotatebox{-90}{
\includegraphics[width=0.39\textwidth,height=0.34\textheight]{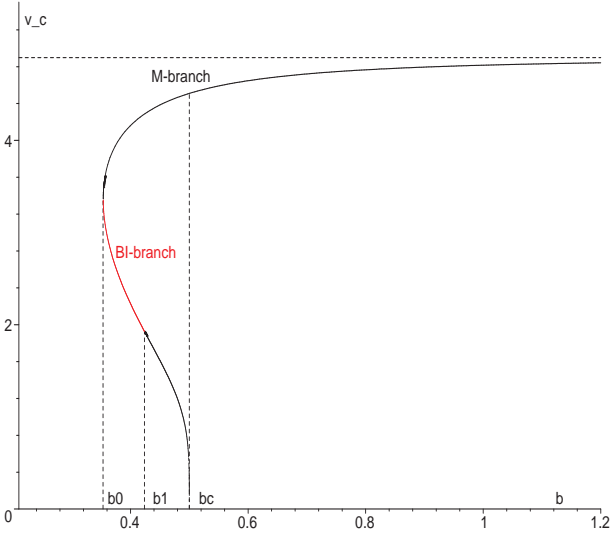}
}
\caption{{\bf Critical volume.}
The dependence of the critical volume $v_c$ on the parameter $b$ is depicted for $Q=1$.
The upper (black) solid line displays the M-branch. It exists for all $b\geq b_0$ and asymptotes to the RN value $v_{RN}=2\sqrt{6} Q$ (upper dashed line).
The critical point of the BI-branch (lower solid line) exists for $b\in(b_0, b_2)$, but has positive pressure only for $b\in (b_0, b_1)$; the corresponding line is displayed by the red solid line.
In this range of parameter $b$ we have two critical points with positive pressure, one for each branch. The critical point of M-branch has higher pressure, temperature, and $\frac{P_c v_c}{T_c}$ ratio (see the following three figures).
}\label{fig:Vc}  
\end{center}
\end{figure}

\begin{figure}
\begin{center}
\rotatebox{-90}
{
\includegraphics[width=0.39\textwidth,height=0.34\textheight]{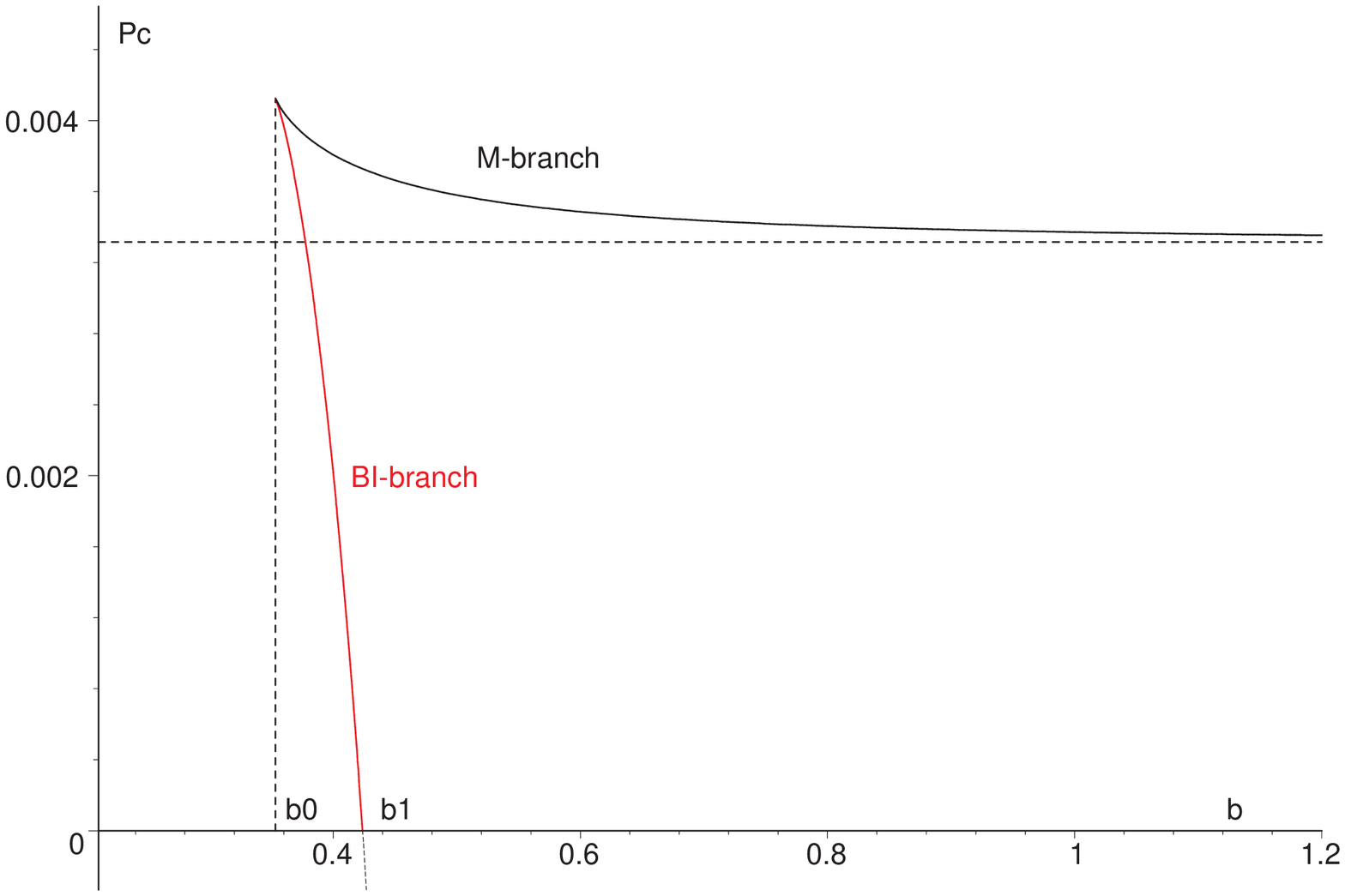}
}
\caption{{\bf Critical pressure.}
The dependence of the critical pressure $P_c$ on the parameter $b$ is depicted for $Q=1$.
The upper (black) solid line displays the M-branch, while the lower solid red line displays the positive pressure part of the BI-branch. The dashed horizontal line stands for the asymptotic RN-AdS value $P_{RN}=1/(96 \pi Q^2)$, towards which the M-branch asymptotes.
}\label{fig:Pc}  
\end{center}
\end{figure}

\begin{figure}
\begin{center}
\rotatebox{-90}
{
\includegraphics[width=0.39\textwidth,height=0.34\textheight]{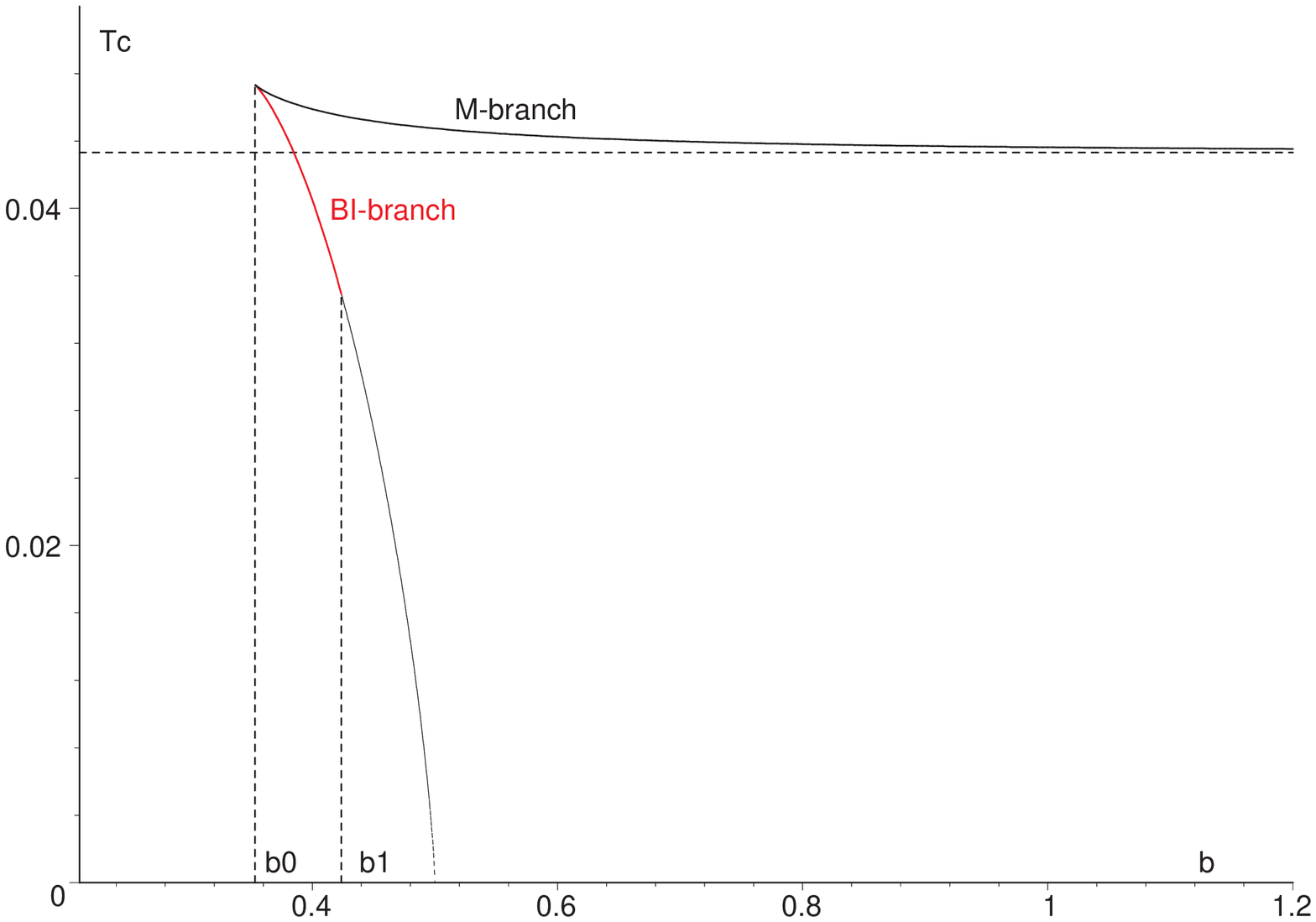}
}
\caption{{\bf Critical temperature.}
The dependence of the critical temperature $T_c$ on the parameter $b$ is depicted for $Q=1$.
The upper (black) solid line displays the M-branch, while the lower solid red line displays the positive pressure part of the BI-branch. The dashed horizontal line stands for the asymptotic RN-AdS value $T_{RN}=\sqrt{6}/(18 \pi Q)$, towards which the M-branch asymptotes.
}\label{fig:Tc}  
\end{center}
\end{figure}

\begin{figure}
\begin{center}
\rotatebox{-90}
{
\includegraphics[width=0.39\textwidth,height=0.34\textheight]{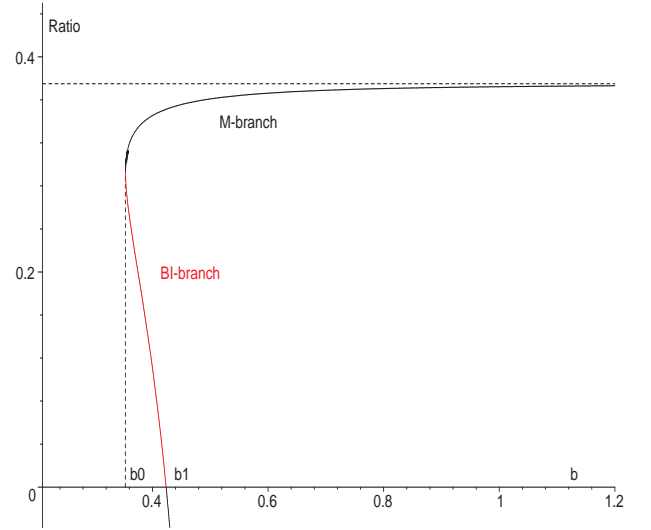}
}
\caption{{\bf Critical ratio.}
The dependence of the critical ratio $\rho_c=\frac{T_c v_c}{P_c}$ on the parameter $b$ is depicted for $Q=1$.
The dashed horizontal is the RN-AdS asymptotic value $3/8$, towards which the M-branch asymptotes.
}\label{fig:rrc}  
\end{center}
\end{figure}

To summarize, a critical point can occur only for $b\geq b_0$. Between $b_0\leq b \leq b_2$ we have two critical points, given by 
\eqref{xk}, with $x=x_{0,1}$, where $x_1$ corresponds to a larger $v_c$. Above $b_2$ there is one critical point, given by $x=x_1$. 
Using a series expansion, it is straightforward to show that for $b\to \infty$, $v_c$ determined from $x_1$ approaches the `specific volume' of the Reissner--N\"ordstrom-AdS black hole. That is why we call the branch determined from $x_1$ the {\em Maxwell (M-branch)}, and the one determined from $x_0$ (which is deeply in the Born--Infeld regime) the {\em BI-branch}. 
Once the critical radius $v_c$ has been determined, 
\be\label{BIvc}
v_c=\left( \frac{1}{x^2}\! -\! \frac{16 Q^2}{b^2} \right)^{\frac{1}{4}}\!\!\!\!,\ \   
x=\left\{ \begin{array}{lll}
x_1\,,&  b\geq b_0& \mbox{(M-branch)} \\
x_0\,, &  b\in(b_0, b_2) &\mbox{(BI-branch)}
\end{array}
\right.
\ee
the critical temperature and the critical pressure are given by 
\ba\label{BITc}
T_c&=&\frac{1-8xQ^2}{\pi v_c}\,,\nonumber\\
P_c&=&\frac{1-16xQ^2}{2\pi v_c^2}-\frac{b^2}{4\pi}\left(1-\frac{1}{v_c^2x}\right)\,.
\ea
Let us note that the BI-branch critical point has negative (unphysical) pressure for 
\be
b\geq b_1=\frac{\sqrt{3+2\sqrt{3}}}{6Q}\approx \frac{0.4237}{Q}\,.
\ee
Excluding this range we conclude that the BI-branch critical point occurs for $b\in (b_0, b_1)$. 
The dependence of the critical quantities on the parameter $b$ as well as the ratio $\rho_c=\frac{P_c v_c}{T_c}$ is illustrated in Figs.~\ref{fig:Vc}--\ref{fig:rrc}.
The values of the quantities for large $b$, described by the M-branch, asymptote to those of the Reissner--N\"ordstrom-AdS black hole, given by formulas \eqref{RNcrit} and
\eqref{RNratio}.  

We finally mention that for $b<b_2$ the black hole has one horizon, i.e., it belongs to the S-type. 
On the other hand for $b>b_2$ (the M-branch) the black hole may be of S or RN type. The transition occurs when 
$M(b, P_c, v_c, Q)=M_m(b,Q)$. Namely, for 
\be
b>b_3\approx \frac{1.72846}{Q}
\ee
the critical point occurs for the RN-type black hole, whereas it occurs for the S-type for $b<b_3$. Obviously, it is the S-type black holes 
which possess more interesting properties.

To study the possible phase transitions in the system, let us now turn to the expression for the Gibbs free energy.

\subsubsection{Gibbs free energy}
In order to find the Gibbs free energy of the system let us calculate its Euclidean action. For a fixed charge $Q$, one considers a surface integral
\cite{MiskovicOlea:2008, MyungEtal:2008} 
\ba
I_s&=&-\frac{1}{8\pi}\int_{\partial M} d^3x\sqrt{h}K\nonumber\\
&&-\frac{1}{4\pi}\int_{\partial M} d^3x \sqrt{h} \frac{n_aF^{ab}A_b}{\sqrt{1+2F/b^2}}\,.
\ea
The first term is the standard Gibbons--Hawking term while the latter term is needed to impose fixed $Q$ as a boundary condition at infinity.
The total action is then given by 
\be
I=I_{BI}+I_s+I_c\,,
\ee
where $I_{EM}$ is given by \eqref{IEM}, and $I_c$ represents the invariant counterterms needed to cure the infrared divergences \cite{EmparanEtal:1999, Mann:1999}.
Slightly more generally, in $d$ dimensions the total action, $I=\beta(M-TS)$, reads   \cite{MiskovicOlea:2008}
\ba
I&=&\frac{\beta \omega_{d-2}}{16\pi}\Bigl[r_+^{d-3}-\frac{r_+^{d-1}}{l^2}\nonumber\\
&&+\frac{4\tilde q^2(d-2)}{(d-1)(d-3)r_+^{d-3}}\,{}_2F_1\!\left(\frac{d\!-\!3}{2d\!-\!4},\frac{1}{2};\frac{3d\!-\!7}{2d\!-\!4};\frac{-\tilde q^2}{b^2r_+^{2d\!-\!4}}\right)\nonumber\ \\
&&-\frac{4b^2r_+^{d-1}}{(d-1)(d-2)}\Bigl(1-\sqrt{1+\frac{\tilde q^2}{b^2r_+^{2d-4}}}\Bigr)\Bigr]\,,
\ea 
with the black hole charge $Q$ given by $Q=\frac{2(d-3)\omega_{d-2}}{8\pi}\tilde q$\,. 
Since the action has been calculated for  fixed $\Lambda$, we associate it with the {\em Gibbs free energy} for  fixed charge (returning back to $d=4$),  
\ba\label{GibbsAdS}
G(T,P)&=&\frac{1}{4}\Bigl[r_+-\frac{8\pi}{3} P r_+^3-\frac{2b^2r_+^3}{3}\left(1-\sqrt{1+\frac{Q^2}{b^2r_+^4}}\right)\nonumber\\
&&+\frac{8Q^2}{3r_+}\,{}_2F_1\!\left(\frac{1}{4},\frac{1}{2};\frac{5}{4};-\frac{Q^2}{b^2r_+^4}\right)\Bigr]\,.
\ea
Here, $r_+$ is understood as a function of pressure and temperature, $r_+=r_+(P,T)$, via equation of state \eqref{BIstate}.
\begin{figure}
\begin{center}
\rotatebox{-90}{
\includegraphics[width=0.39\textwidth,height=0.34\textheight]{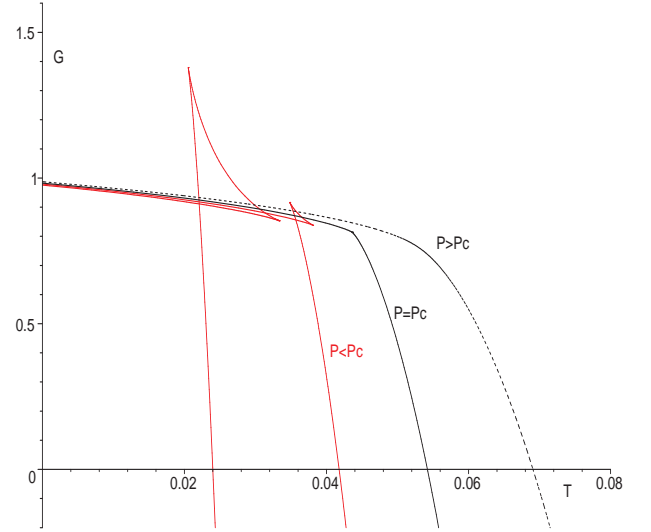}
}
\caption{{\bf Gibbs free energy for $b>b_2$.}
The behaviour of the Gibbs free energy is depicted as a function of temperature for fixed pressure. For $b>b_2$ the behaviour is reminiscent of that of the RN-AdS black hole, i.e., there is one critical point and the corresponding first order phase transition
between small and large black holes for $T<T_c$. We have set $Q=1$ and 
$b=1$.
}\label{fig:GTBIa} 
\end{center}
\end{figure} 
\begin{figure}
\begin{center}
\rotatebox{-90}{
\includegraphics[width=0.39\textwidth,height=0.34\textheight]{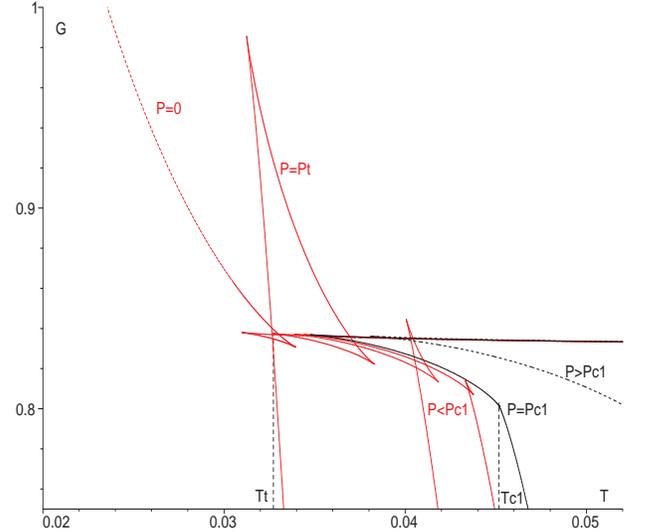}
}
\caption{{\bf Gibbs free energy for $b\in (b_1, b_2)$.}
In this range of parameters one has one physical (with positive pressure) critical point and the corresponding first oder phase transition between small and large black holes, occurring for $T\in (T_t, T_{c_1})$ and $P\in(P_t, P_{c_1})$. Note that contrary to the case $b>b_2$, small black holes ($r_+\to 0$) now correspond to high temperature, $T\to \infty$. There also exists a certain range of pressures, $P\in(P_t, P_z)$, and temperatures, $T\in (T_t, T_z)$, for which the global minimum of $G$ is discontinuous (see Fig.~\ref{fig:zerothorder}) and, aside from the first order phase transition, there are two phases of intermediate and small black holes separated by a finite jump in $G$. For $T<T_t$ only one phase of large black holes exists, see $P-T$ diagram in Fig.~\ref{fig:PTbMagnified}. We have set $Q=1$ and $b=0.45$.
}\label{fig:GTBIb}  
\end{center}
\end{figure} 
\begin{figure}
\begin{center}
\rotatebox{-90}{
\includegraphics[width=0.39\textwidth,height=0.34\textheight]{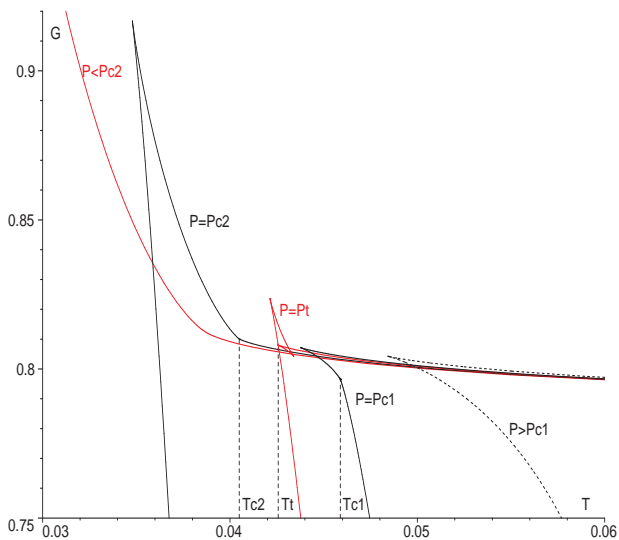}
}
\caption{{\bf Gibbs free energy for $b\in(b_0,b_1)$.}
In this range one has two critical points with positive pressure. However, only the one at $T=T_{c_1}$ corresponds to the first order phase transition between small and large black holes. The other does not globally minimize the Gibbs free energy and hence is unphysical. As $b$ decreases these critical points move closer each other and finally merge and disappear completely for $b=b_0$. Similar to the previous figure in between $(T_t, T_z)$ there is a `zeroth order phase transition' between intermediate and small black holes.
We have set $Q=1$ and $b=0.4$.
}\label{fig:GTBIc}  
\end{center}
\end{figure} 
\begin{figure}
\begin{center}
\rotatebox{-90}{
\includegraphics[width=0.39\textwidth,height=0.34\textheight]{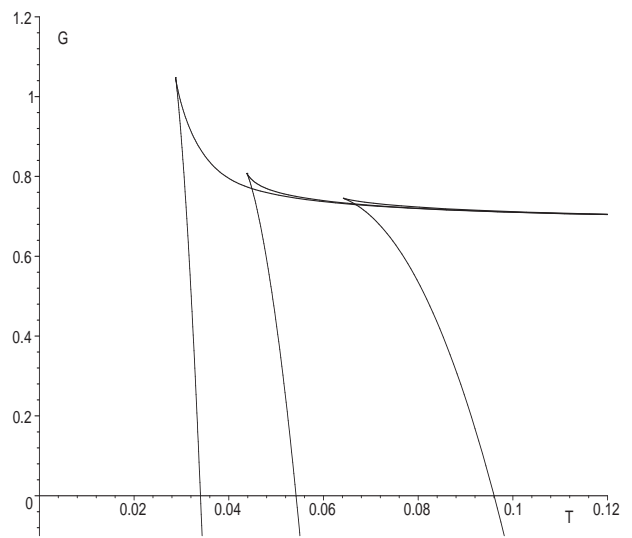}
}
\caption{{\bf Gibbs free energy for $b<b_0$.}
For $b<b_0$, similar to the Schwarzschild-AdS case, there is no first order phase transition in the system. We have set
$Q=1$ and $b=0.3$.
}\label{fig:GTBId}  
\end{center}
\end{figure} 

The behaviour of $G$ is depicted in Figs.~\ref{fig:GTBIa}--\ref{fig:GTBId}. It depends crucially 
on the value of parameter $b$. Namely, for $b>b_2$ it is similar to that of the RN-AdS black hole, i.e., there is one critical point and the corresponding first order phase transition
between small and large black holes for $T<T_c$. The $P-T$ diagram remains qualitatively the same as for the 4D RN-AdS black hole and is almost identical to Fig.~\ref{fig:PT4D}.

\begin{figure}
\begin{center}
\rotatebox{-90}{
\includegraphics[width=0.39\textwidth,height=0.34\textheight]{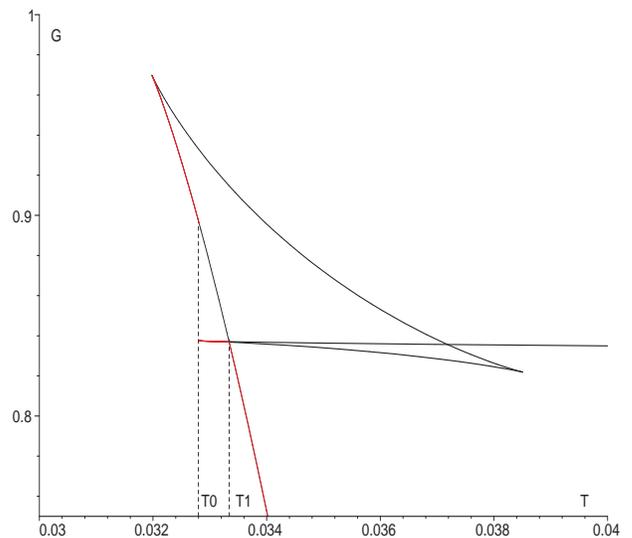}
}
\caption{{\bf Zeroth order phase transition.}
The global minimum of the Gibbs free energy  (displayed by the black solid line) is highlighted by the red thick line. We have set $Q=1$, $b=0.45$ and 
$P=0.46 \times Pc_1\approx 1.6854 \times 10^{-3}\in (P_t, P_z)$. Obviously, there is a first order phase transition between small and large black holes occurring at $T=T_1\approx 3.3340\times 10^{-2}$. There is also a discontinuity in the global minimum of $G$ at $T=T_0\approx 3.2799\times 10^{-2}$. This separates the intermediate size black holes from the small black holes. We refer to this as a `zeroth-order phase transition'. This type of behaviour is characteristic for all $b\in (b_0, b_2)$ and  $P\in (P_t, P_z)$.
}\label{fig:zerothorder}  
\end{center}
\end{figure}

\begin{figure}
\begin{center}
\rotatebox{-90}{
\includegraphics[width=0.39\textwidth,height=0.34\textheight]{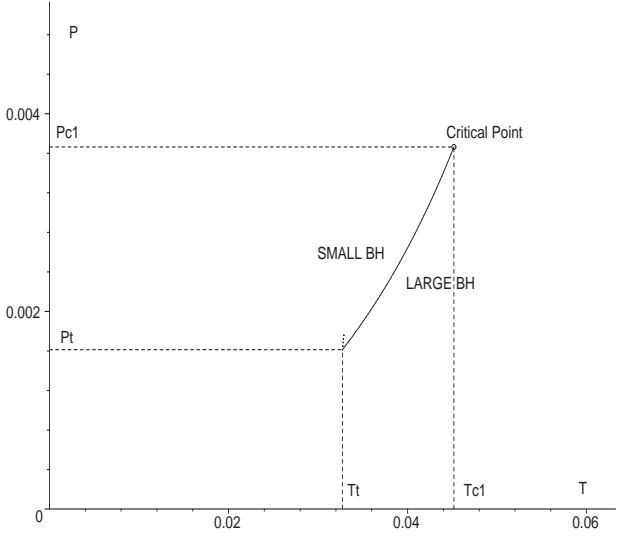}
}
\caption{{\bf $P-T$ diagram for $b\in(b_1, b_2)$.}
The coexistence line of the first order phase transition between small and large black holes is depicted by a thick solid black line for $Q=1$ and $b=0.45$. It initiates from the critical point $(P_{c_1}, T_{c_1})$ and terminates at $(P_t, T_t)$. 
There is also a `zeroth order phase transition' (dotted line) discussed in more detail in the next figure.
}\label{fig:PTb}  
\end{center}
\end{figure}

\begin{figure}
\begin{center}
\rotatebox{-90}{
\includegraphics[width=0.39\textwidth,height=0.34\textheight]{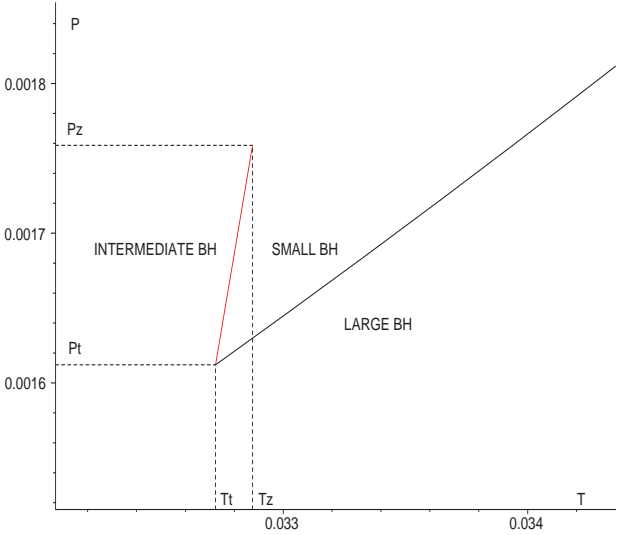}
}
\caption{{\bf $P-T$ diagram magnified: zeroth order phase transition.}
The first order phase transition between small and large black holes is displayed by thick solid black line. The red solid line describes the `coexistence line' of small and intermediate black holes, separated by a finite gap in $G$, indicating the zeroth order phase transition. It commences from $(T_z, P_z)$ and terminates at $(P_t, T_t)$.
}\label{fig:PTbMagnified}  
\end{center}
\end{figure}

For $b\in (b_1, b_2)$ one has one physical (with positive pressure) critical point and the corresponding first order phase transition between small and large black holes. This phase transition occurs for $T<T_{c_1}$ and terminates at $T=T_t$. 
There also exists a certain range of temperatures, $T\in (T_t, T_z)$, for which the global minimum of $G$ is discontinuous, see Fig.~\ref{fig:zerothorder}. In this range of temperatures two separate branches of intermediate size and small size black holes co-exist. They are separated by a finite jump in $G$.  Although there is no real phase transition between them, we refer (for simplicity) to this phenomenon as {\em `zeroth-order phase transition'}.
For $T<T_t$ only one phase of large black holes exists: see the $P-T$ diagram in Fig. \ref{fig:PTb} 
and its magnification in Fig.~\ref{fig:PTbMagnified}. 

For $b\in (b_0, b_1$) there exist two critical points with positive pressure. However, only the one at $T=T_{c_1}$ corresponds to the first order phase transition between small and large black holes. The other, with lower temperature, lower pressure, smaller radius $r_+$, and smaller mass, does not globally minimize the Gibbs free energy and hence is unphysical. As $b$ decreases the two critical points move closer to each other and finally merge and disappear completely for $b=b_0$. Similar to the case $b\in (b_1, b_2)$, there is a `zeroth order phase transition' between intermediate and small black holes. For $T<T_t$ only a phase of large black holes exists. The $P-T$ diagram is very similar to the case when $b\in(b_1, b_2)$.

Finally, for $b<b_0$ there is no first order phase transition in the system and the behaviour is like that of the Schwarzschild--AdS black hole.

\subsubsection{Critical exponents}
In the preceding discussion, we have learned that for all $b>b_0$ there is a first order (small--large)-black hole phase transition in the system,  which terminates at the critical point described by $(v_c, T_c, P_c)$, given by Eqs. \eqref{BIvc} and \eqref{BITc}. This critical point occurs for the M-branch, i.e., it is described by $x=x_1$, \eqref{xk}. [The other possible critical point belonging to the BI-branch is unphysical since the corresponding phase of the system does not globally minimize the Gibbs free energy.] For large $b$ the critical values expand as  
\ba
v_c&=&2\sqrt{6}Q-\frac{7\sqrt{6}}{216}\frac{1}{Qb^2}+O\left(\frac{1}{b^4}\right)\,,\nonumber\\
T_c&=&\frac{\sqrt{6}}{18Q\pi}+\frac{\sqrt{6}}{2592 \pi Q^3 b^2}+O\left(\frac{1}{b^4}\right)\,,\nonumber\\
P_c&=&\frac{1}{96\pi Q^2}+\frac{7}{41472\pi  Q^4 b^2}+O\left(\frac{1}{b^4}\right)\,,
\ea
and asymptote to those of the RN-AdS black hole \eqref{RNcrit}. Exact values (for any $b>b_0$) are  given by 
\eqref{BIvc} and \eqref{BITc} and are displayed in Figs.~\ref{fig:Vc}--\ref{fig:Tc} (M-branch). The critical ratio $\rho_c=\frac{P_c v_c}{T_c}$ is displayed in Fig.~\ref{fig:rrc}. For large $b$ it expands as 
\be
\rho_c=\frac{3}{8}-\frac{1}{384 Q^2b^2}+O\left(\frac{1}{b^4}\right)\,,
\ee  
and asymptotes to the value $3/8$, characteristic for the Van der Waals fluid, \eqref{VdWcrit}, or the RN-AdS black hole, \eqref{RNratio}.

Let us now study the vicinity of this critical point by calculating its critical exponents. The entropy $S$, as function of $T$ and $V$ is given by 
\be
S=\pi r_+^2=\left(\frac{9\pi V}{16}\right)^{1/3} \,.
\ee
It follows that $C_V=0$ and hence the critical exponent $\alpha=0$. To calculate other critical exponents we employ the equation of state \eqref{BIstateV}. We introduce the quantities $p$, $\omega$ and 
$t$, given by \eqref{reduced} and \eqref{omegat} with $z=3$, and series expand the law of corresponding states,  which is again of the form \eqref{general}, to obtain 
\be\label{pBI}
p=1+At-Bt\omega-C\omega^3+O(t\omega^2, \omega^4)\,,
\ee 
where 
\be
A=\frac{1}{\rho_c}\,,\quad B=\frac{1}{3\rho_c}\,,
\ee
and $C$, given by a relatively complicated expression, whose exact form is not really important. For simplicity, we give the 
large $b$ expansion 
\be
C=\frac{4}{81}-\frac{7}{2916 b^2Q^2}+O\left(\frac{1}{b^4}\right)\,,
\ee 
which obviously asymptotes to the RN-AdS value $4/81$. 
Following the discussion in Sec.~II.C, it is now obvious that 
the critical exponents $\alpha$, $\beta$, $\gamma$, and $\delta$ take the same values as in the (higher-dimensional) 
Maxwell case, 
\be\label{BIexp}
\alpha=0,\quad \beta=\frac{1}{2}\,,\quad \gamma=1\,, \quad \delta=3\,.
\ee 
So we have shown that even for the Born--Infeld black holes, obeying highly nonlinear electromagnetic field equations, we obtain the same critical exponents as in the linear Maxwell case.

Let us finally mention that the derived critical exponents \eqref{BIexp} differ from those of recent papers 
\cite{BanerjeeEtal:2010, BanerjeeRoychowdhury:2012, BanerjeeRoychowdhury:2012b, LalaRoychowdhury:2011}.  
The authors therein study the critical behaviour of the Born--Infeld-AdS black hole system in non-extended phase space and 
obtain various critical exponents (different from ours) which do not agree with the mean field theory prediction. The reason for this is that they do not really study the vicinity of the (Van der Waals) critical point, but rather the behaviour close to the two points characterized by horizon radii $r_1$ and $r_2$ where the specific heat at constant charge $C_Q$ diverges. Such points, however, correspond to the (Van der Waals) critical point only in the limit when these radii coincide, i.e., $r_1=r_2=r_c$, in which case the analysis in these papers is no longer valid. In all other cases the two studied points do not correspond to the global minimum of the Gibbs free energy and hence are unphysical. Hence the analysis in these papers is not correct. 
We expect that a proper analysis,  e.g., similar to \cite{NiuEtal:2011}, would lead to the same value of the critical exponents as derived in our paper.

\subsection{Vacuum polarization}
Let us finally briefly return to the new quantity ${\cal B}$, 
\be\label{Bnew}
{\cal B}=\frac{2}{3}br_+^3\Bigl(1-\sqrt{1\!+\!\frac{Q^2}{b^2r_+^4}}\Bigr)
+\frac{Q^2}{3br_+}\,{}_2F_1\!\left(\frac{1}{4},\frac{1}{2}; \frac{5}{4};-\frac{Q^2}{b^2r_+^4}\right)\!,
\ee
which was recently computed as a conjugate quantity to the Born--Infeld parameter $b$ \cite{Huan:2010}.
An interesting question is a physical interpretation of this quantity. Since $b$ has units of the electric intensity and $M$ units of energy, we can see from formula \eqref{Smarr} that ${\cal B}$ has units of electric polarization (or more precisely a polarization per unit volume). Hence we refer to ${\cal B}$ {\em `Born--Infeld vacuum polarization'}.  This quantity is necessarily non-vanishing in Born--Infeld space times and is essential for consistency of the Smarr relation  \eqref{Smarr}.

\begin{figure}
\begin{center}
\rotatebox{-90}{
\includegraphics[width=0.39\textwidth,height=0.34\textheight]{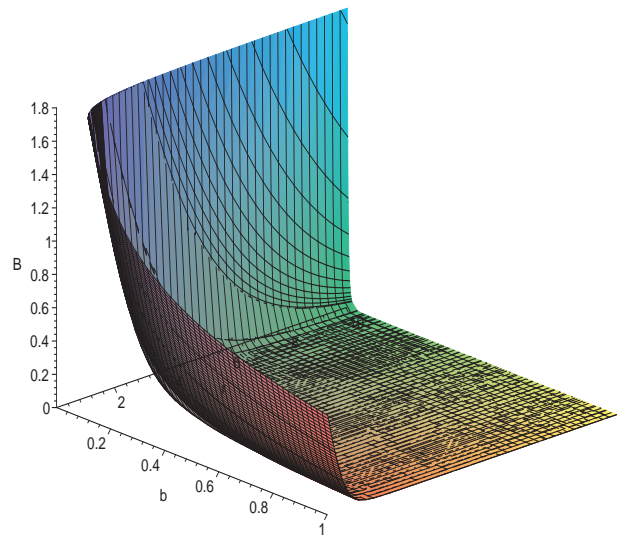}
}
\caption{{\bf Vacuum polarization ${\cal B}$.}
The behaviour of ${\cal B}$ as a function of $b$ and $r_+$ is displayed. Apparently, as we approach the linear Maxwell case ($b$ increases) or we are far from the source ($r_+$ increases) ${\cal B}$ rapidly decreases.  For $r_+=0$ and non-trivial $b$ we have a finite value of ${\cal B}$, as further confirmed by analysis and the following figure.
}\label{fig:Bbr}  
\end{center}
\end{figure}

\begin{figure}
\begin{center}
\rotatebox{-90}{
\includegraphics[width=0.39\textwidth,height=0.34\textheight]{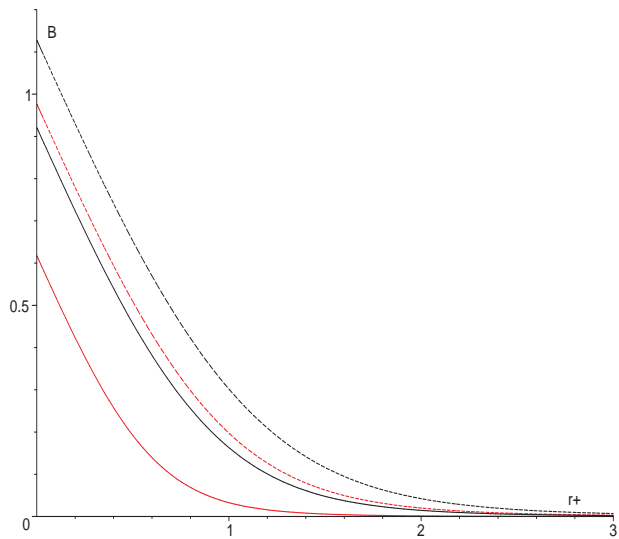}
}
\caption{{\bf Vacuum polarization ${\cal B}$ for fixed $b$.}
The behaviour of ${\cal B}$ as a function of $r_+$ is displayed for $b=1, 0.45, 0.4$ and $0.3$ (from bottom up). Note that for $r_+=0$, ${\cal B}$ reaches the finite value.
}\label{fig:Brplus}  
\end{center}
\end{figure}

The behaviour of ${\cal B}$ as a function of $r_+$ and $b$ is depicted in Figs.~\ref{fig:Bbr} and \ref{fig:Brplus}. 
 Apparently, as we approach the linear Maxwell case ($b$ increases) or we are further away from the source ($r_+$ increases) ${\cal B}$ rapidly decreases  to zero.  On the hand, for $r_+ \to 0$ and a non-trivial value of $b$, the quantity ${\cal B}$ reaches a finite positive value, see Fig.~\ref{fig:Brplus}. 

\begin{figure}
\begin{center}
\rotatebox{-90}{
\includegraphics[width=0.39\textwidth,height=0.34\textheight]{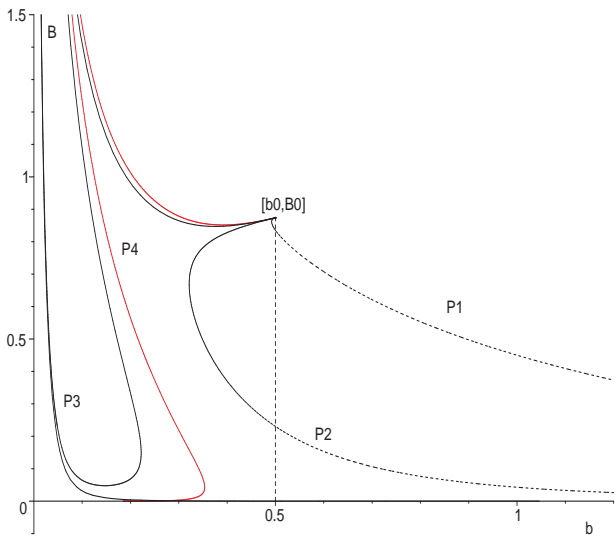}
}
\caption{{\bf ${\cal B}-b$ diagram.}
Isobars in $({\cal B}, b)$-plane are depicted for various choices of $P$ and fixed $T=3T_{RN}=\sqrt{6}/(6\pi Q)$. All isobars 
 emerge from the point $[b_0, {\cal B}_0]$ which corresponds to small black holes, $r_+\to 0$.
The subsequent behaviour as $r_+$ increases depends on the value of $P$. Namely, the displayed curves correspond to $P/P_{RN}=500, 20, 4$ and $0.2$, with $P_{RN}=1/(96\pi Q^2)$. We have set $Q=1$. Note that the $P_3$ curve cusps and meets the ${\cal B}=0$ axis near $b=0.25$.
}\label{fig:BbT1}  
\end{center}
\end{figure} 
\begin{figure}
\begin{center}
\rotatebox{-90}{
\includegraphics[width=0.39\textwidth,height=0.34\textheight]{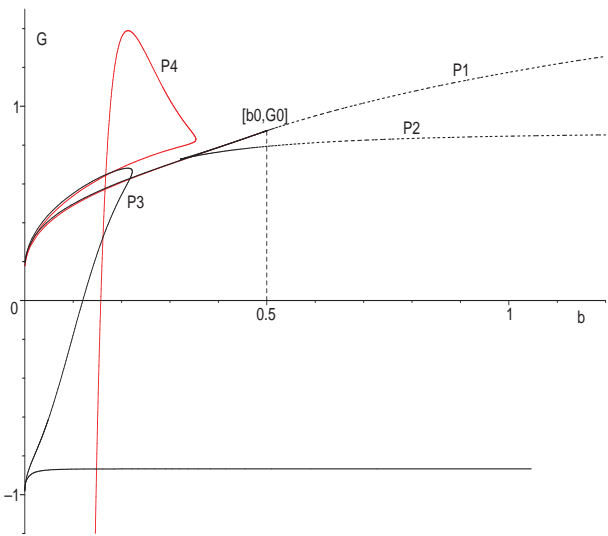}
}
\caption{{\bf Gibbs free energy as function of $b$.}
$G$ is depicted as function of $b$ for various values of $P$ and fixed $T=3T_{RN}$. Similar to the previous figure, all curves 
start from the point $[b_0, G_0]$ and the subsequent behaviour depends on the ratio $P/P_c$. We have set $Q=1$ and the values of $P$ as in previous figure.  
}\label{fig:GbT1}  
\end{center}
\end{figure} 

Using Eq. \eqref{BIstate} we can rewrite $r_+$ as function of $P, T, b$ and $Q$, thereby obtaining from \eqref{Bnew}  a `new' equation of state
\be
{\cal B}={\cal B}(b, Q, P, T)
\ee
  in the extended phase space (including $b$ and ${\cal B}$ variables).  Similarly the expression \eqref{GibbsAdS} is now understood as $G=G(b, Q, P, T)$.  It then makes sense to construct a `$({\cal B}-b)$-diagram', drawing the isobars (isotherms) for fixed $Q$ and $T$ ($P$). We display this in Fig.~\ref{fig:BbT1}, which  demonstrates   interesting behaviour that is further confirmed by the $G-b$ diagram displayed in Fig.~\ref{fig:GbT1}. We leave the analysis of these diagrams (and a possible critical behaviour therein) as possible future directions for research.


\section{Summary}
We have investigated the thermodynamic behaviour of a broad range of charged and rotating black holes in the context of an extended thermodynamic phase space.  For Einstein--Maxwell theory this phase space includes the conjugate pressure/volume quantities, where the former is proportional to the cosmological constant and the latter is proportional to a term that would be
the Newtonian geometric volume of the black hole.  These relationships hold in any dimension, and are necessary for the Smarr relation to be valid.    We find in all dimensions $d\geq 4$ that an analogy with the Van der Walls liquid--gas system holds, provided we identify the specific volume in the
Van der Waals equation with the radius of the event horizon (and not the thermodynamic volume) multiplied by an appropriate power of the Planck length. However only in $d=4$ does the critical ratio $\rho_c =\frac{P_c v_c}{T_c}$, \eqref{PcvcTc}, equal its corresponding
Van der Waals value of $3/8$. The critical exponents also coincide with those of the Van der Waals system, due to a universal behaviour, Eq. (\ref{general}), in which
the pressure is equal to a term proportional to the temperature divided by the specific volume plus a function $f(v)$ only of the specific volume.
This situation is a consequence of the form of the metric function in Eq. (\ref{metfunction}) and holds for all the black holes we consider in this paper, including  neutral slowly rotating black holes in $d=4$. The only exceptions are charged or rotating black holes in $d=3$, for which no critical behaviour exists due to the particular form of $f(v)$. 

We also examined the thermodynamic behaviour of Born--Infeld black holes.  Here we found that the phase space needs to be further extended, to include not only the pressure/volume terms from the Einstein-Maxwell case but also the conjugate pair $({\cal B},b)$. 
Inclusion of this latter pair is necessary to obtain consistency of the Smarr relation.  The (recently computed \cite{Huan:2010}) former quantity, ${\cal B}$, we refer to as the Born--Infeld polarization because of its dimensionality.  While we have analyzed to some
extent the behaviour of this quantity in Figs.~\ref{fig:BbT1} and \ref{fig:GbT1}, it is clear that its full physical meaning remains to be understood.  We find for Born--Infeld black holes two interesting types---the RN-type and the S-type---in which the horizon
structure is quite distinct.   While for sufficiently large $b$ the behaviour of the Gibbs free energy is the same as for the Einstein-Maxwell case, once $b<b_2 = 1/(2Q)$ a second critical point emerges, and the thermodynamic behaviour markedly changes (though the critical exponents do not change).  The most intriguing feature we observe is that the Gibbs free energy becomes discontinuous in a certain temperature range, indicating a new kind of phase transition between small and intermediate sized black holes.  We call this a zeroth-order phase transition, in keeping with the nomenclature that a first-order phase transition is one in which the Gibbs-free energy is continuous but not differentiable whereas for a second order transition both the Gibbs free energy and its first derivatives are continuous.   While the physics of this discontinuity remains to be understood, we expect that the intermediate/small transition will result in a burst of radiation from the black hole as a consequence of the sudden drop in the Gibbs free energy.
 
 None of the examples we considered are outside of the  the mean field theory prediction for critical exponents.  Evidently the inclusion of non-linear electrodynamic effects are not sufficient to go beyond mean field theory, and so it will be necessary to include quantum gravitational corrections to do so.  A recent attempt along these lines \cite{ZhangLi:2012} posits an ansatz for the quantum corrected
 black hole temperature for Reissner--N\"ordstrom black holes and then proceeds to investigate   how the critical exponents will correspondingly be modified.  However the critical exponents are inappropriately identified, and so the question remains open.  We expect to return to this and other interesting subjects in the future.

\vspace{0.2cm}

\section*{Acknowledgments}
We would like to thank D. Dalidovich for illuminating discussions and C. Pope for useful comments at the early stage of this work. We also like to thank D. Kastor for reading the manuscript. This work was supported in part by the Natural Sciences and Engineering Research Council of Canada.


\appendix

\section{Critical exponents}
{\em Critical exponents} describe the behaviour of physical quantities near the critical point. It is believed that they are {\em universal}, i.e., they do not depend on the details of the physical system, though they may depend on the dimension of the system and/or the range of interactions. In $D\geq 4$ dimensions they can be calculated by using the mean field (Landau's) theory, e.g., \cite{Goldenfeld:1992}.

Let us consider a fluid characterized by the critical point which occurs at critical temperature $T_c$, critical specific volume $v_c$, and critical pressure $P_c$.  We further define
\be
t=\frac{T-T_c}{T_c}=\tau-1\,,\quad \phi=\frac{v-v_c}{v_c}=\nu-1\,. 
\ee
Then the critical exponents $\alpha, \beta, \gamma, \delta$ are defined as follows:
\begin{itemize}
\item
Exponent $\alpha$ governs the behaviour of the specific heat at constant volume, 
\be
C_v=T \frac{\partial S}{\partial T}\Big|_{v}\propto |t|^{-\alpha}\,.
\ee
\item Exponent $\beta$ describes the behaviour of the {\em order parameter} $\eta=v_g-v_l$ (the difference of the volume of the gas $v_g$ phase and the volume of the liquid phase $v_l$)  on the given isotherm
\be
\eta =v_g-v_l\propto |t|^\beta\,.
\ee
\item 
Exponent $\gamma$ determines the behaviour of the {\em isothermal compressibility} $\kappa_T$, 
\be
\kappa_T=-\frac{1}{v}\frac{\partial v}{\partial P}\Big|_T\propto |t|^{-\gamma}\,.
\ee
\item Exponent $\delta$ governs the following behaviour on the critical isotherm $T=T_c$:  
\be
|P-P_c|\propto |v-v_c|^\delta\,.
\ee
\end{itemize}

In particular, for the fluid described by the Van der Waals equation in any dimension $d>3$ (see appendix B)
\be\label{vdwA}
\left(P+\frac{a}{v^2}\right)(v-b)=T\,,
\ee
where we have set the Boltzmann constant $k=1$, and $a>0$, $b>0$ are constants describing the size of molecules and their interactions. The critical point occurs at
\be\label{abc}
T_c=\frac{8a}{27b}\,,\quad v_c=3b\,,\quad P_c=\frac{a}{27b^2}\,,
\ee
giving the universal critical ratio 
\be\label{VdWcrit}
\rho_c=\frac{P_c v_c}{T_c}=\frac{3}{8}\,.
\ee
Using these quantities and defining
\be\label{pnutau}
p=\frac{P}{P_c}\,,\quad \nu=\frac{v}{v_c}\,,\quad \tau =\frac{T}{T_c}\,,
\ee
we can rewrite the Van der Waals equation in the universal (independent of $a$ and $b$) {\em law of corresponding states} 
\be\label{states}
8\tau=(3\nu-1)\left(p+\frac{3}{\nu^2}\right)\,.
\ee
The critical exponents are then calculated to be (see, e.g., \cite{KubiznakMann:2012})
\be
\alpha=0\,,\quad \beta=\frac{1}{2}\,,\quad \gamma=1\,,\quad \delta=3\,,
\ee
which are the values predicted by the mean field theory.

Similarly, for a spherical charged AdS black hole in 4D we have the following equation of state \cite{KubiznakMann:2012}:
\be
P=\frac{T}{v}-\frac{1}{2\pi v^2}+\frac{2Q^2}{\pi v^4}\,.
\ee
Here, $P$ stands for the pressure associated with the cosmological constant, $P=-\frac{1}{8\pi}\Lambda$, $T$ is the the black hole temperature,
$Q$ its charge, and $v$ is the corresponding `fluid specific volume' which can be associated with the horizon radius $r_+$ as 
\be
v=2l_P^2r_+\,.
\ee
This equation admits a critical point which occurs at 
\be\label{RNcrit}
T_c=\frac{\sqrt{6}}{18\pi Q}\,,\quad v_c=2\sqrt{6} Q\,,\quad P_c=\frac{1}{96\pi Q^2}\,.
\ee
Similar to the Van der Waals equation, these quantities satisfy
\be\label{RNratio}
\frac{P_c v_c}{T_c}=\frac{3}{8}\,,
\ee
independently of the charge of the black hole. Using further definitions \eqref{pnutau}, 
we get the universal (independent of $Q$) `law of corresponding states'  
\be\label{statesR}
8\tau=3\nu\left(p+\frac{2}{\nu^2}\right)-\frac{1}{\nu^3}\,.
\ee
The critical exponents are calculated to be 
\be
\alpha=0\,,\quad \beta=\frac{1}{2}\,,\quad \gamma=1\,,\quad \delta=3\,,
\ee
and coincide with those of the Van der Waals fluid.

\section{Van der Waals equation in higher dimensions: heuristic derivation}
Let us repeat the statistical mechanics derivation of an equation of state of a weakly interacting gas.
In $d$ number of spacetime dimensions, the partition function of $N$ interacting particles with positions and momenta 
$\bf{r}_i\,, \bf{p}_i$\,, $i=1,\dots, N$ is given by 
\ba
Z&=&\int d{\bf r}_1\dots d{\bf r}_N d{\bf p}_1\dots d{\bf p}_N e^{-\beta H}\,,\nonumber\\
&&H=\sum_i \frac{{\bf p}_i^2}{2m}+\sum_{i<j}W_{ij}\,,
\ea
with $\beta=kT$, and $W_{ij}$ being the two-particle interactions.
Integrating over momenta, while using the formula $\int_{-\infty}^\infty dp \exp(-\beta\frac{p^2}{2m})=\sqrt{\frac{2\pi m}{\beta}}$, and expanding the interacting part of the exponential ($\beta>>\sum W_{ij}$), we get 
\ba
Z&=&\Bigl(\frac{2\pi m}{\beta}\Bigr)^{\frac{N(d-1)}{2}}
\int d{\bf r}_1\dots d{\bf r}_N \Bigl(1-\beta \sum_{i<j} W_{ij}\Bigr)\,\nonumber\\
&=&\Bigl(\frac{2\pi m}{\beta}\Bigr)^{\frac{N(d-1)}{2}}
V^N\Bigl(1-\beta\frac{N(N-1)}{2V^2} \int\!\! d{\bf r}_1 d{\bf r}_2W_{12}\Bigr)\,,\nonumber\\
\ea
with $V$ being the volume of the gas. 
Going to the centre of mass system, we can re-write 
the last integral as
\be
\int W_{12}d{\bf r}_1 d {\bf r}_2=-2Va\,,\quad a=-\frac{1}{2}\int W(|\vec{r}|) d{\bf r}\,.
\ee
Using large $N$ approximation we finally have 
\be
Z=\Bigl(\frac{2\pi m}{\beta}\Bigr)^{\frac{N(d-1)}{2}}
V^N\Bigl(1+\frac{\beta N^2a}{V}\Bigr)\,.
\ee
The free energy is then defined as
$F=-kT \log Z$, which taking $a$ small, the expansion becomes 
\be
F=-\frac{d-1}{2}kTN\log(2\pi m kT)-kTN\log V-\frac{N^2a}{V}+O(a^2)\,.
\ee
This thermodynamic potential obeys $dF=-PdV-S dT+\mu dN$.
The equation of state then reads $P=-\frac{\partial F}{\partial V}$, which in terms of the specific volume $v=V/N$ reads
\be\label{state1}
\left(P+\frac{a}{v^2}\right)v=kT\,.
\ee
One can also calculate the entropy from $S=-\frac{\partial F}{\partial T}$, or calculate the specific heats for constant volume or pressure 
\ba
C_v&=&T\frac{\partial S}{\partial T}\Bigr|_V=\frac{k(d-1)N}{2}\,,\nonumber\\
C_p&=&T\frac{\partial S}{\partial T}\Bigr|_P=\frac{k(d+1)N}{2}+\frac{2PN}{T^2 k}a+O(a^2)\,.
\ea
Hence we find 
\be
\kappa=\frac{C_p}{C_v}=\frac{d+1}{d-1}+\frac{4P}{k^2T^2(d-1)}a+O(a^2)\,,
\ee
generalizing {\em Myer's law}, $\kappa=5/3+O(a)$ to higher dimensions. 
As a final step we mention that it is quite natural to generalize \eqref{state1} by taking into account the 
finite volume of gas molecules $b$ and hence `deriving' the Van der Waals equation in higher dimensions:
\be\label{VdWstate2}
\left(P+\frac{a}{v^2}\right)(v-b)=kT\,.
\ee
Obviously, such an equation of state inherits all the properties of the standard Van der Waals equation.

\section{Hypergeometric identities}
Hypergeometric function ${}_pF_q$ is defined for $a_1,\dots, a_p$, $b_1,\dots,b_p,z\in {\mathbb C}$ as follows
\be
{}_pF_q(a_1,\dots, a_p;b_1,\dots,b_q;z)=\sum_{j=0}^\infty
\frac{(a_1)_j\dots (a_p)_j}{(b_1)_j\dots (b_q)_j}\frac{z^j}{j!}\,,
\ee
where the Pochhammer symbol $(\mu)_j$ is defined as $(\mu)_0=1$ and $(\mu)_j=\mu(\mu+1)\dots (\mu+j-1)$ for  $j=1,2,\dots$

In what follows we concentrate on ${}_2F_1$ needed in the main text, for which the series converges for $|z|<1$ (and $|z|=1$ if $\Re(c-a-b)>0$).
We have the following asymptotic expansion as $z\to 0$: 
\be
\,{}_2F_1\!\left(\frac{1}{4},\frac{1}{2};\frac{5}{4};-z\right)=1-\frac{z}{10}+\frac{1}{24}z^2+O(z^3)\,, 
\ee
A known integral representation for $\Re(c)>0$, $\Re(b)>0$, (and any $z$ modulo a cut along the real axis from 1 to infinity), reads
\be
{}_2F_1(a,b;c;z)=\frac{\Gamma(c)}{\Gamma(b)\Gamma(c-b)}\int_0^1dt \frac{t^{b-1}(1-t)^{c-b-1}}{(1-zt)^a}\,,
\ee
which for $a=1/2$ and $c=b+1$ reduces to 
\be
{}_2F_1\left(\frac{1}{2},b;b+1;-z\right)=b \int_0^1dt \frac{t^{b-1}}{\sqrt{1+zt}}\,.
\ee
Using this formula we have 
\be
\int_r^\infty \frac{dx}{\sqrt{x^4+r_0^4}}=\frac{1}{r}\ {}_2F_1\left(\frac{1}{4},\frac{1}{2};\frac{5}{4};-\frac{r_0^4}{r^4}\right)\,.
\ee
Integrating by parts we get 
\ba\label{C6}
\int_r^\infty \!\!\!&dx& \bigl(\sqrt{x^4+r_0^4}-x^2\bigr)\nonumber\\
\!\!&=&\frac{2r_0^4}{3r}{}_2F_1\left(\frac{1}{4},\frac{1}{2};\frac{5}{4};-\frac{r_0^4}{r^4}\right)
\!+\!\frac{r^3}{3}\Bigl(1\!-\!\sqrt{1+\frac{r_0^4}{r^4}}\Bigr)\,,\quad\ 
\ea
which is used in the main text. We also find
\be
\int_0^\infty\!\!\frac{dx}{\sqrt{x^4+r_0^4}}=\frac{1}{4r_0\sqrt{\pi}}\,\Gamma\Bigl(\frac{1}{4}\Bigr)^2\,.
\ee
Hence, by expanding around $r=0$ we get 
\ba
\int_r^\infty\!\!\!\frac{dx}{\sqrt{x^4+r_0^4}}&=&\int_0^\infty\frac{dx}{\sqrt{x^4+r_0^4}}-\int_0^r\frac{dx}{\sqrt{x^4+r_0^4}}\nonumber\\
&=&\frac{1}{4r_0\sqrt{\pi}}\,\Gamma\Bigl(\frac{1}{4}\Bigr)^2-\frac{r}{r_0^2}+O(r^2)\,.\quad
\ea
Integrating further by parts we get the following expansion around $r=0$, used in the main text:
\ba\label{expansionzero}
\int_r^\infty\!\!\!\!\!&&\!\!\!\!\!\bigl(\sqrt{x^4+r_0^4}-x^2\bigr)dx
\nonumber\\
&=&\frac{1}{3}r^3-\frac{1}{3}r\sqrt{r^4+r_0^4}+\frac{2}{3}r_0^4\int_r^\infty \!\!\!\frac{dx}{\sqrt{x^4+r_0^4}}\nonumber\\
&=&\frac{r_0^3}{6\sqrt{\pi}} \Gamma\Bigl(\frac{1}{4}\Bigr)^2-r_0^2r+O(r^2)\,.
\ea
Finally, using the formula for the derivative of ${}_2F_1$,
\be\label{A1}
\frac{d}{dz}{}_2F_1(a,b;c;z)=\frac{ab}{c}{}_2F_1(a+1,b+1;c+1;z)\,,
\ee
the recurrence relation
\ba
{}_2F_1(a+&&1,b;c;z)-{}_2F_1(a,b;c;z)\nonumber\\
&&=\frac{bz}{c}{}_2F_1(a+1,b+1,c+1;z)\,,
\ea
and the fact that 
\be
{}_2F_1(5/4,1/2;5/4;-z)=\frac{1}{\sqrt{1+z}}\,,
\ee
we find the following two important relations used in the main text: 
\be\label{C13}
\frac{d}{dz}\,{}_2F_1\!\left(\frac{1}{4},\frac{1}{2};\frac{5}{4};-z\!\right)=
\frac{a}{z}\!\left[\frac{1}{\sqrt{1\!+\!z}}\!-{}_2F_1\!\left(\frac{1}{4},\frac{1}{2};\frac{5}{4};-z\right)\right]\!.
\ee
\be\label{A5}
\,{}_2F_1\!\left(\frac{5}{4},\frac{3}{2};\frac{9}{4};-z\right)=
\frac{c}{bz}\left[\frac{1}{\sqrt{1\!+\!z}}\!-{}_2F_1\!\left(\frac{1}{4},\frac{1}{2};\frac{5}{4};-z\right)\right]\!.
\ee




\providecommand{\href}[2]{#2}\begingroup\raggedright\endgroup

\end{document}